\documentclass[final,5p,times,twocolumn,numbers,sort&compress]{elsarticle}

\usepackage{graphicx}
\usepackage{amsmath}
\usepackage{subfigure}
\usepackage{wrapfig}
\usepackage{epsfig}
\usepackage{afterpage,float}
\usepackage{slashbox}
\newcommand\fverb{\setbox\fverbbox=\hbox\bgroup\verb}
\newcommand\fverbdo{\egroup\medskip\noindent%
			\fbox{\unhbox\fverbbox}\ }
\newcommand\fverbit{\egroup\item[\fbox{\unhbox\fverbbox}]}
\newbox\fverbbox

\newcommand{\be}{\begin{equation}}
\newcommand{\ee}{\end{equation}}
\newcommand{\bea}{\begin{eqnarray}}
\newcommand{\eea}{\end{eqnarray}}
\newcommand{\nn}{\nonumber}

\def\half{{\textstyle{1\over2}}}

\def\MeV{\mathop{\rm MeV}\nolimits}
\def\GeV{\mathop{\rm GeV}\nolimits}
\def\Tr{{\sf Tr}}
\def\Re{{\sf Re}}
\def\Im{{\sf Im}}
\def\Det{{\sf Det}}
\def\bar{\overline}
\def\hat{\widehat}
\def\tilde{\widetilde}

\def\half{{\scriptstyle \raise.15ex\hbox{${1\over2}$}}}

\newcommand{\beq}{\begin{equation}}
\newcommand{\eeq}{\end{equation}}

\newcommand{\real}{\relax{\rm I\kern-.18em R}}

\def\vek#1{{\bf #1}}
\def\overbar{\overline}
\def\vv{{\bf V}}
\def\rd{{\rm d}}
\def\mco{\multicolumn}

\usepackage{graphicx}
\usepackage{afterpage,float}

\usepackage{amssymb}
\journal{Physics Letters B}

\begin{document}

\begin{frontmatter}

\def\MeV{\mathop{\rm MeV}\nolimits}
\def\GeV{\mathop{\rm GeV}\nolimits}
\def\Tr{{\sf Tr}}
\def\Re{{\sf Re}}
\def\Im{{\sf Im}}
\def\Det{{\sf Det}}
\def\bar{\overline}
\def\hat{\widehat}
\def\tilde{\widetilde}

\def\half{{\scriptstyle \raise.15ex\hbox{${1\over2}$}}}

\newcommand{\Cite}[1]{$\,$\cite{#1}}

\def\vek#1{{\bf #1}}
\def\overbar{\overline}
\def\vv{{\bf V}}
\def\rd{{\rm d}}
\def\mco{\multicolumn}

\title{Twelve massless flavors and three colors below the conformal window}
 
 \author[wupi,budapest]{Zolt\'{a}n Fodor}
%\ead{fodor@bodri.elte.hu}
 
 \author[uop]{Kieran Holland}
% \ead{kholland@pacific.edu}
 
 \author[ucsd]{Julius Kuti\corref{cor1}}
 \ead{jkuti@ucsd.edu}
 
 \author[budapest]{D\'{a}niel N\'{o}gr\'{a}di}
% \ead{nogradi@lorentz.leidenuniv.nl}
 
 \author[wup]{Chris Schroeder}
% \ead{crs@physics.ucsd.edu}

 \cortext[cor1]{Corresponding author}
% \cortext[cor2]{Principal corresponding author}
 %\fntext[fn1]{This is the specimen author footnote.}
 %\fntext[fn2]{Another author footnote, but a little more longer.}
 %\fntext[fn3]{Yet another author footnote. Indeed, you can have
%    any number of author footnotes.}
 
 \address[wupi]{Department of Physics, University of Wuppertal, 
Gaussstrasse 20, D-42119, Germany\\
J\"ulich Supercomputing Center, Forschungszentrum, 
        J\"ulich, D-52425 J\"ulich, Germany}
 \address[uop]{Department of Physics, University of the Pacific, 
3601 Pacific Ave, Stockton CA 95211, USA}
 \address[ucsd]{Department of Physics 0319, University of California, San Diego, 
9500 Gilman Drive, La Jolla, CA 92093, USA}
\address[wup]{Department of Physics, University of Wuppertal, 
Gaussstrasse 20, D-42119, Germany}
\address[budapest]{Institute for Theoretical Physics, E\"otv\"os University, 
        H-1117 Budapest, Hungary}

\begin{abstract}We report new results for a frequently discussed
gauge theory with twelve fermion flavors  
in the fundamental  representation
of the SU(3) color gauge group. The model, controversial with respect to its conformality,
is important in non-perturbative studies searching for a viable composite 
Higgs mechanism Beyond the Standard Model (BSM). 
To resolve the controversy, we subject the model to opposite hypotheses inside and outside of
the conformal window. In the first hypothesis we test  chiral symmetry breaking ($\chi{\rm SB}$) with
its Goldstone spectrum, $F_\pi$,  the $\chi{\rm SB}$ condensate, 
and several composite hadron states as the fermion mass is varied
in a limited range with our best effort to control finite volume effects and extrapolation to the massless chiral limit.
Supporting results for $\chi{\rm SB}$ from the running coupling based on the force between static sources 
and some preliminary evidence for the finite temperature transition are also presented.
In the second test for the alternate hypothesis we probe conformal behavior driven by a single anomalous mass dimension
under the assumption of unbroken chiral symmetry. Our  results show a very low level of confidence
in the conformal scenario. 
Staggered lattice fermions with stout-suppressed taste breaking are used throughout the simulations.
\end{abstract}

\begin{keyword}
lattice simulations, electroweak sector, technicolor, conformal

%% keywords here, in the form: keyword \sep keyword

%% PACS codes here, in the form: \PACS code \sep code

%% MSC codes here, in the form: \MSC code \sep code
%% or \MSC[2008] code \sep code (2000 is the default)

\end{keyword}

\end{frontmatter}

\section{Introduction}

New physics at the Large Hadron Collider could be discovered in the form of some new strongly-interacting gauge
theory with a composite Higgs mechanism, an idea which was outside experimental 
reach when it was first introduced as an attractive BSM scenario~\cite{Weinberg:1979bn,Susskind:1978ms,Dimopoulos:1979es,
Eichten:1979ah,Farhi:1980xs,Holdom:1984sk,Yamawaki:1985zg,Appelquist:1987fc,Miransky:1996pd}. 
The original framework has  been expanded by new
explorations of the multi-dimensional theory space of nearly conformal gauge theories~\cite{Caswell:1974gg,Banks:1981nn,Appelquist:2003hn, Sannino:2004qp,
Dietrich:2005jn,Luty:2004ye,Dietrich:2006cm,Kurachi:2006ej}
where systematic non-perturbative lattice studies 
play a very important role.
New experimental results at the Tevatron~\cite{Aaltonen:2011mk}, boldly interpreted as Technicolor~\cite{Eichten:2011sh},
will further stimulate lattice efforts to provide a well-controlled
theoretical framework.
Interesting models require the theory to be very
close to, but below, the conformal window, with a running coupling
which is almost constant over a large energy range. 
The non-perturbative knowledge of the critical flavor $N^{crit}_f$ separating the two phases is essential and 
this has generated much interest and many new lattice
studies~\cite{Fodor:2009wk,Fodor:2011tw,Appelquist:2007hu,Appelquist:2009ty,Appelquist:2009ka,Deuzeman:2008sc,
Deuzeman:2009mh,Hasenfratz:2009ea,Hasenfratz:2010fi,Jin:2009mc,Jin:2010vm,Catterall:2007yx,Catterall:2008qk,Hietanen:2008mr,
Hietanen:2009az,DelDebbio:2010hx,Bursa:2010xn,Shamir:2008pb,DeGrand:2010na,Kogut:2010cz,Sinclair:2010be,Bilgici:2009kh,
Itou:2010we,Yamada:2009nt,Hayakawa:2010yn}.  

We report new studies of an important and frequently discussed gauge theory
with twelve fermion flavors  
in the fundamental  representation of the SU(3) color gauge group.  With $N_f=12$ being close to the critical flavor number,
the model has attracted a great deal 
of attention in the lattice community and off-lattice as well.
To establish the chiral properties of a gauge theory close to the conformal window is notoriously difficult.
If the chiral symmetry is broken, the fundamental parameter $F$ of the chiral Lagrangian has to be small in 
lattice units $a$  to control cut-off effects. Since the chiral expansion has terms with powers 
of  $N_f M^2_\pi/16\pi^2 F^2$, reaching the chiral regime with a large number of fermion flavors is
particularly difficult. 
The range of $aM_\pi$ values where 
leading chiral logs can be identified unambiguously will require simulations in very large volumes 
which are not in the scope of this study.
We will make a case in this report that 
qualitatively different expectations inside and outside the conformal window allow
tests of the two mutually exclusive hypotheses without reaching down to the chiral logs at very small pion masses. 

Below the conformal window,
chiral symmetry is broken at zero fermion mass with a gap in the composite hadron spectrum 
except for the associated massless Goldstone multiplet. 
The analytic form of the chiral Lagrangian as a function of the
fermion mass can be used to detect chiral log corrections, or to differentiate from conformal exponents in the transitional
region before the chiral logs are reached at low enough Goldstone pion masses.  
Approximations to gauge theories with  $\chi{\rm SB}$, 
like their effective Nambu-Jona-Lasinio 
description in the large N limit, are consistent with this analysis. 
In sharp contrast, the spectrum inside the conformal window is gapless in all channels 
in the chiral limit and the scale dependence of physical quantities,
like the fermion mass dependence of 
composite operators and their correlators,  is governed by the single 
critical exponent $\gamma$. 

The two competing hypotheses can be  tested in search for chiral properties
of gauge theories. 
%It is important to emphasize a rather powerful argument here.
There is a fundamental difference between the two hypotheses as implied by their respective
spectra.  $\chi{\rm SB}$ creates 
a fundamental scale $F$ in the theory separated from the composite hadron scale with 
its  residual baryon gap in the chiral limit.
The pion mass can be varied from the $\chi{\rm SB}$ scale $F$ to the hadron scale with a transition
from the chiral log regime to a regime without chiral analysis. The conformal phase has no
intrinsic scale.
With  $\chi{\rm SB}$ this is expected to lead to fermion mass dependence of the spectrum 
in the chiral log regime, or above it, quite different from
the conformal behavior which is very tightly constrained near the chiral limit 
of the spectrum with a single critical exponent
$\gamma$ in the absence of any intrinsic scale.
In a regime where lattice cutoff effects are negligible, this difference 
should be sufficient for  tests whether the chiral
loop expansion is reached, or not, on the low scale $F$.

\begin{table*}[!ht]
\caption{Measured masses and  $F_\pi$ with  the three largest volumes in the $m=0.01-0.02$ range and the largest volume for $m>0.02$. Asterisks indicate $L_s=32$ when different from the spatial volume of the second column. $M_{pnuc}$ is the mass of the nucleon's parity partner.}
\vskip 0.05in
\resizebox{\textwidth}{!}{
\centering
\begin{tabular}{|l|l|l|l|l|l|l|l|l|l|l|l|}
\hline\hline
mass & lattice &$M_\pi$ & $F_\pi$& $M_{i5}$&$M_{sc}$ &$M_{ij}$ & $M_{nuc}$&$M_{pnuc}$ &$M_{Higgs}$&$M_{rho}$ &$M_{A1}$\\
\hline                                      
\hline  0.0100& $48^3\times  96$ &0.1647(23)& 0.02474(49) & 0.1650(13)        &0.16437(95)&0.1657(10)      &0.3066(69)         &0.3051(81)         &0.247(13)  &0.1992(28)       &0.2569(83)          \\ 
\hline  0.0100& $40^3\times  80$ &0.1819(28)& 0.02382(39) & 0.1842(29)        &0.1835(35)&0.1844(44)        &0.3553(93)         &0.352(16)           &0.2143(81)&0.2166(73)       &0.237(12)            \\ 
\hline  0.0100& $32^3\times  64$ &0.2195(35)& 0.02234(46) & 0.2171(31)        &0.194(10)  &0.195(11)          &0.386(16)           &0.387(22)           &0.2162(53)&0.239(19)         & 0.246(21)           \\ 
\hline  0.0150& $48^3\times  96$ &0.2140(14)& 0.03153(51) & 0.2167(16)        &0.2165(17)&0.2185(18)        &0.3902(67)         &0.3881(84)         &0.296(13)  &0.2506(33)       & 0.3245(87)         \\ 
\hline  0.0150& $40^3\times  80$ &0.2200(23)& 0.03167(53) & 0.2210(21)        &0.2218(30)&0.2239(34)        &0.4095(84)         &0.411(10)           &0.291(11)  &0.2574(36)        & 0.327(14)           \\ 
\hline  0.0150& $32^3\times  64$ &0.2322(34)& 0.03168(64) & 0.2319(11)        &0.2318(17)&0.2341(16)        &0.4387(60)         &0.4333(84)         &0.2847(33)&0.2699(41)        &0.324(16)            \\ 
\hline  0.0200& $40^3\times  80$ &0.2615(17)& 0.03934(56) & 0.2736(22)$^*$&0.2651(8)  &0.2766(42)$^*$&0.4673(62)         &0.4699(66)         &0.330(17)  &0.3049(28)        & 0.361(32)           \\ 
\hline  0.0250& $32^3\times  64$ &0.3098(18)& 0.04762(53) & 0.3179(17)        &0.3183(18)&0.3231(20)        &0.563(12)           &0.563(14)           &0.4137(88)&0.3683(19)        & 0.469(14)           \\ 
\hline  0.0275& $24^3\times  48$ &0.3348(29)& 0.05218(85) & 0.3430(18)        &0.3425(25)&0.3471(26)        &0.609(21)           &0.628(23)           &0.460(16)  &0.4050(69)        & 0.523(34)           \\ 
\hline  0.0300& $24^3\times  48$ &0.3576(15)& 0.0561(11)   & 0.3578(15)$^*$&0.3726(29)&0.3790(40)        &0.640(12)$^*$   &0.633(16)$^*$   &0.470(15)  &0.4160(26)$^*$& 0.5222(90)$^*$ \\ 
\hline  0.0325& $24^3\times  48$ &0.3699(66)& 0.0588(15)   & 0.3790(34)        &0.3814(62)&0.3879(62)        &0.680(18)           &0.686(26)           &0.500(21)  &0.4481(39)        &0.548(31)             \\ 
\hline  0.0350& $24^3\times  48$ &0.3927(17)& 0.06422(57) & 0.4065(18)        &0.4074(19)&0.4149(26)        &0.703(28)           &0.741(20)           &0.538(30)  &0.4725(64)        & 0.669(65)            \\ 
\hline 
\end{tabular} }
\label{table:1}
\end{table*}

In Sections 2 and 3 we present new results for the gauge  model with twelve fermions in the fundamental representation.
A new kind of gauge dynamics 
is expected to appear at intermediate distances with 
walking gauge coupling or a conformal fixed point. This has remained controversial with recent efforts 
from five lattice groups~\cite{Fodor:2009wk, Fodor:2011tw, Appelquist:2007hu, Appelquist:2009ty, 
Appelquist:2009ka, Deuzeman:2008sc, Deuzeman:2009mh, Hasenfratz:2009ea, Hasenfratz:2010fi, Jin:2009mc, Jin:2010vm}.
We have made considerable progress to resolve the controversies including tests of the chiral
condensate and the spectrum which favor chiral symmetry breaking with unusual chiral dynamics. 
Applying the $\chi{\rm SB}$ 
 hypothesis to the Goldstone pion,  $F_\pi$, the chiral condensate, and the stable nucleon state collectively leads to
a result of ${\rm \chi^2/dof=1.22}$ representing a high level of confidence. 
With the conformal hypothesis  we find ${\rm \chi^2/dof=8.79}$ representing a very low level of confidence.
Applying a global analysis to all states we measured, the contrasting behavior is somewhat less dramatic but remains significant.
New results on the running coupling from the static force  and our 
simulation of a rapid finite temperature transition in Polyakov loop distributions,  reported elsewhere
and  expected  in association with 
$\chi{\rm SB}$ and its restoration, provide further support for our findings.
In Section 4 we will briefly summarize our conclusions with outlook for future work.

%Although our new results should not be viewed as definitive, they are quite powerful.
%In all fits we were on a fine-grained lattice in the pion mass range $aM_{\pi}=0.16-0.39$ 
%and rho mass range $M_{\rho}=0.2-0.47$. In contrast, a previous study~\cite{Deuzeman:2009mh}
%which reported conformal behavior was in the $aM_{\pi}=0.35-0.67$ range and rho mass range $M_{\rho}=0.39-0.77$.

We have used the tree-level Symanzik-improved gauge action for all simulations in this paper.
The conventional $\beta=6/g^2$ lattice gauge coupling is defined as the overall
factor in front of the well-known terms of the Symanzik lattice action.  Its value is $\beta=2.2$ for all simulations
reported here for the $N_f=12$ model. The link
variables in the staggered fermion matrix were exponentially smeared with  two
stout steps~\cite{Morningstar:2003gk}; the precise definition of the action is given in~\cite{Aoki:2005vt}.  
The RHMC and HMC algorithms were deployed in all runs.
For the molecular dynamics we made use of multiple time scales \cite{Urbach:2005ji} and the
Omelyan integrator \cite{Takaishi:2005tz}.
Our error analysis of  hadron masses which combines systematic 
and statistical effects follows the frequentist histogram approach 
of the Budapest-Marseille-Wuppertal collaboration~\cite{ Durr:2010aw}. 
%The 68 percent width of the histogram of all effective mass fits, weighted with their Q-values for
%fits starting at $N_t/4$ on the plateaux, was taken as the error in each channel. 
The topological charge was monitored
in the simulations with frequent changes observed.

\section{Tests of the $\chi{\rm SB}$ hypothesis}

The chiral Lagrangian for the Goldstone spectrum separated from the massive composite 
scale of hadrons exhibits, order by order, the well-known analytic form of  powers in the fermion mass $m$  
with non-analytic chiral  log corrections 
generated from pion loops close enough to the chiral limit~\cite{Gasser:1983yg}. 
The exact  functions $F_\pi(m)$ and $M_\pi(m)$ will be approximated by an analytic form in powers of
$m$ which is expected to hold over a limited $m$ range when the Goldstone pion is in transition from
 the chiral log regime  toward the composite hadron scale. Although this procedure has some
 inherent uncertainty without the chiral logs directly reached in simulations, its sharp contrast with the 
 non-analytic fermion mass dependence of the conformal hypothesis, governed by the single exponent $\gamma$, is
 sufficient to differentiate the two hypotheses.
 
First, we will illustrate the fitting procedure with results on the Goldstone spectrum, $F_\pi$, and the chiral 
condensate. This will be extended  
to the nucleon and some other composite hadron channels to probe parity degeneracy and residual masses in the chiral limit.
\subsection{Goldstone spectrum and $ F_\pi$ from  $\chi{\rm SB}$}
\begin{figure*}[!htb]
\begin{center}
\begin{tabular}{ccc}
\includegraphics[height=4.8cm]{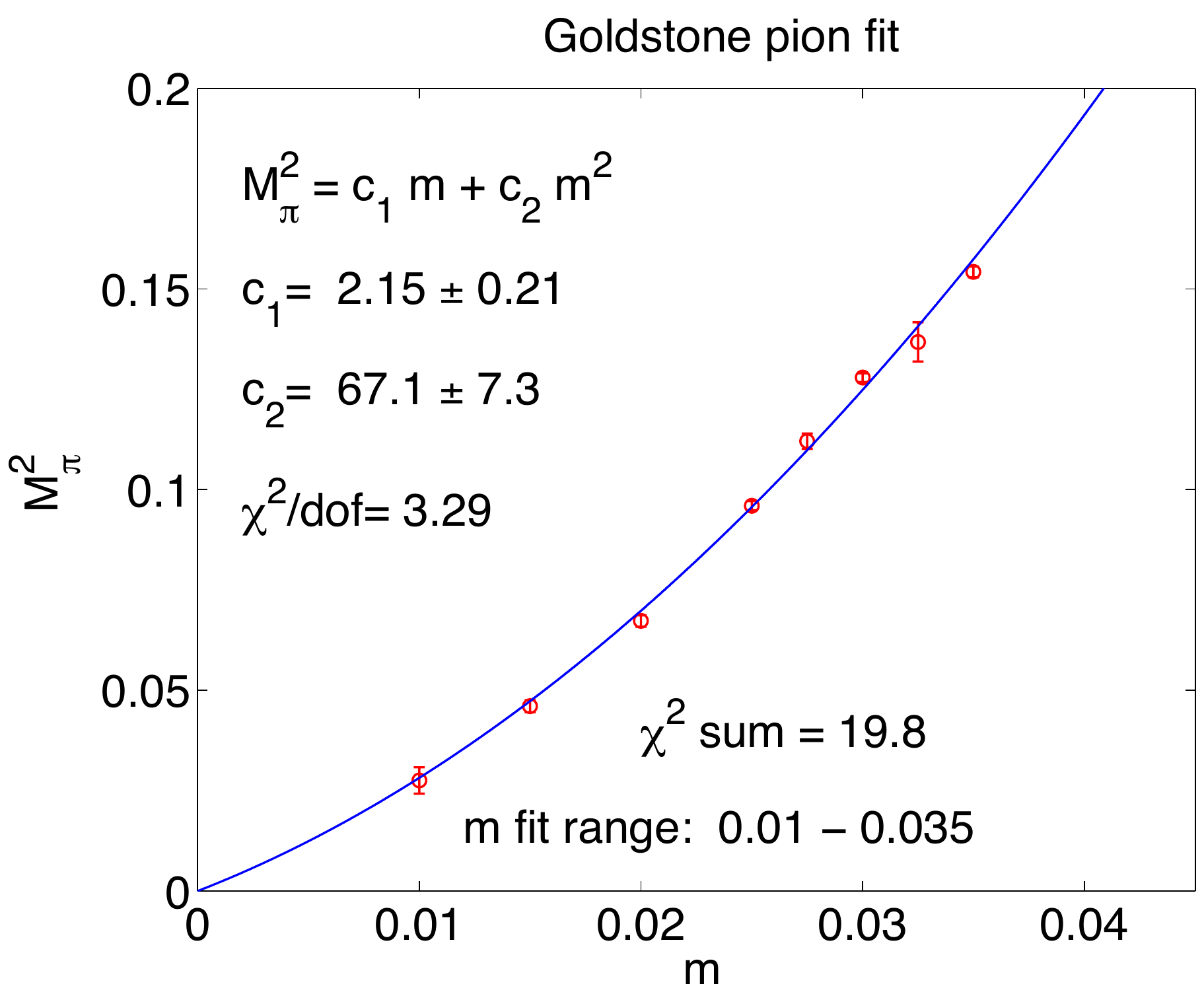}&
\includegraphics[height=4.8cm]{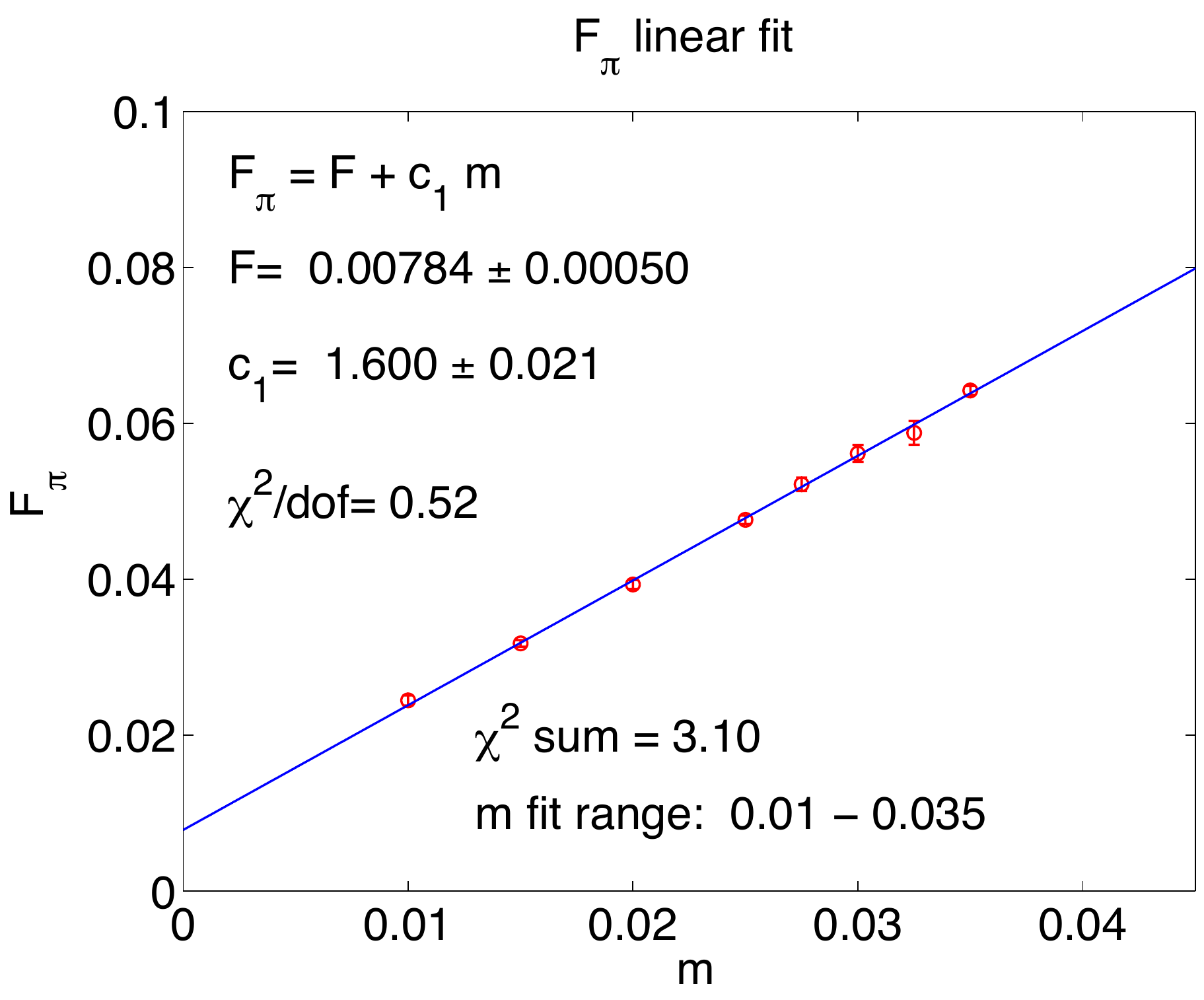}
\includegraphics[height=4.8cm]{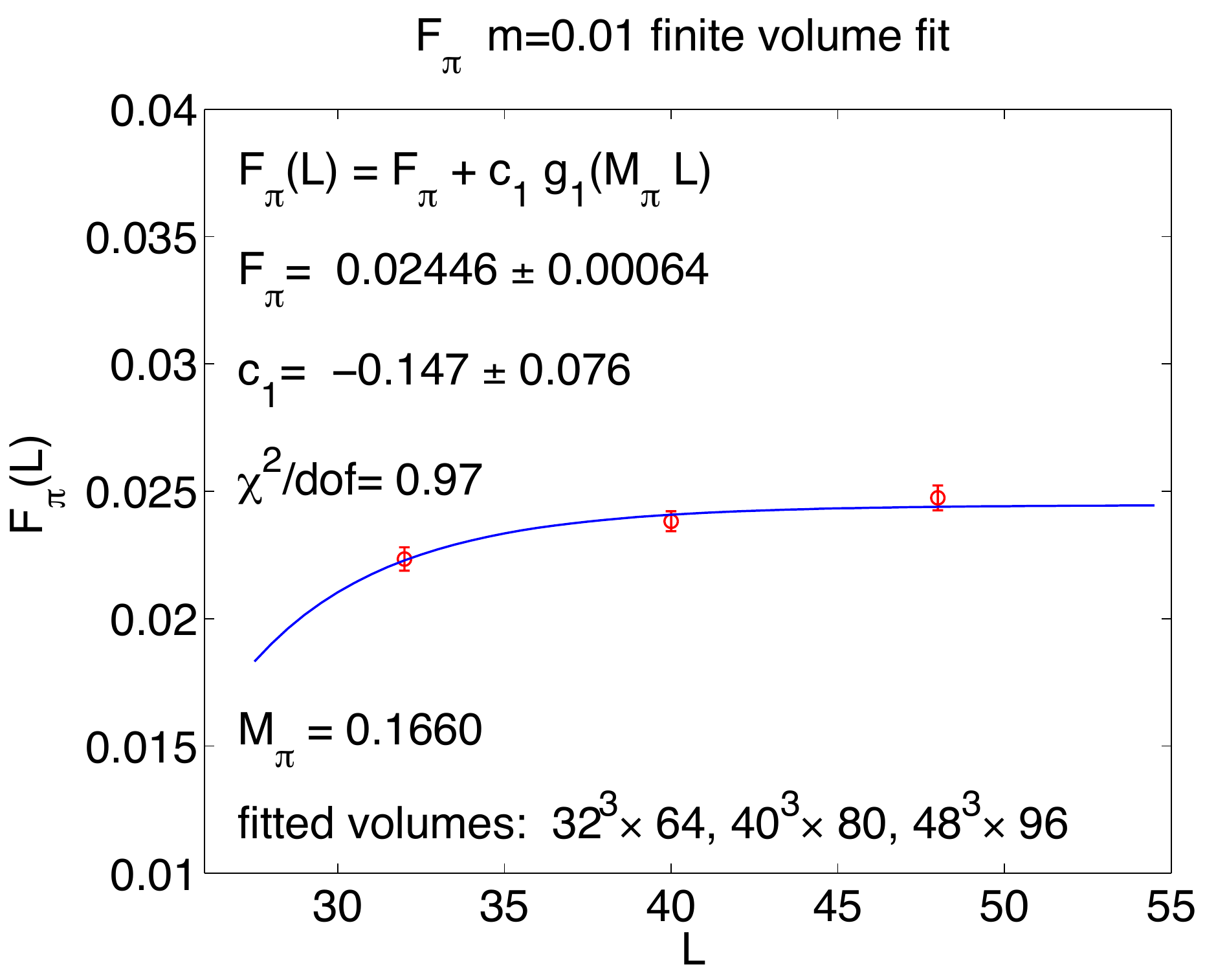}
\end{tabular}
\caption{{\footnotesize The Goldstone pion and $ F_\pi$ from  
chiral symmetry breaking are shown with the fitting procedure described in the text.
A representative finite volume fit is also shown. The infinite volume limit of $M_\pi$ was used in fits to $F_\pi$ and other composite
hadron states, like the nucleon.}}
\end{center}
\label{fig:Fpi-Pion}
\vskip -0.2in
\end{figure*}

\begin{figure*}[!htb]
\begin{center}
\begin{tabular}{ccc}
\includegraphics[height=4.6cm]{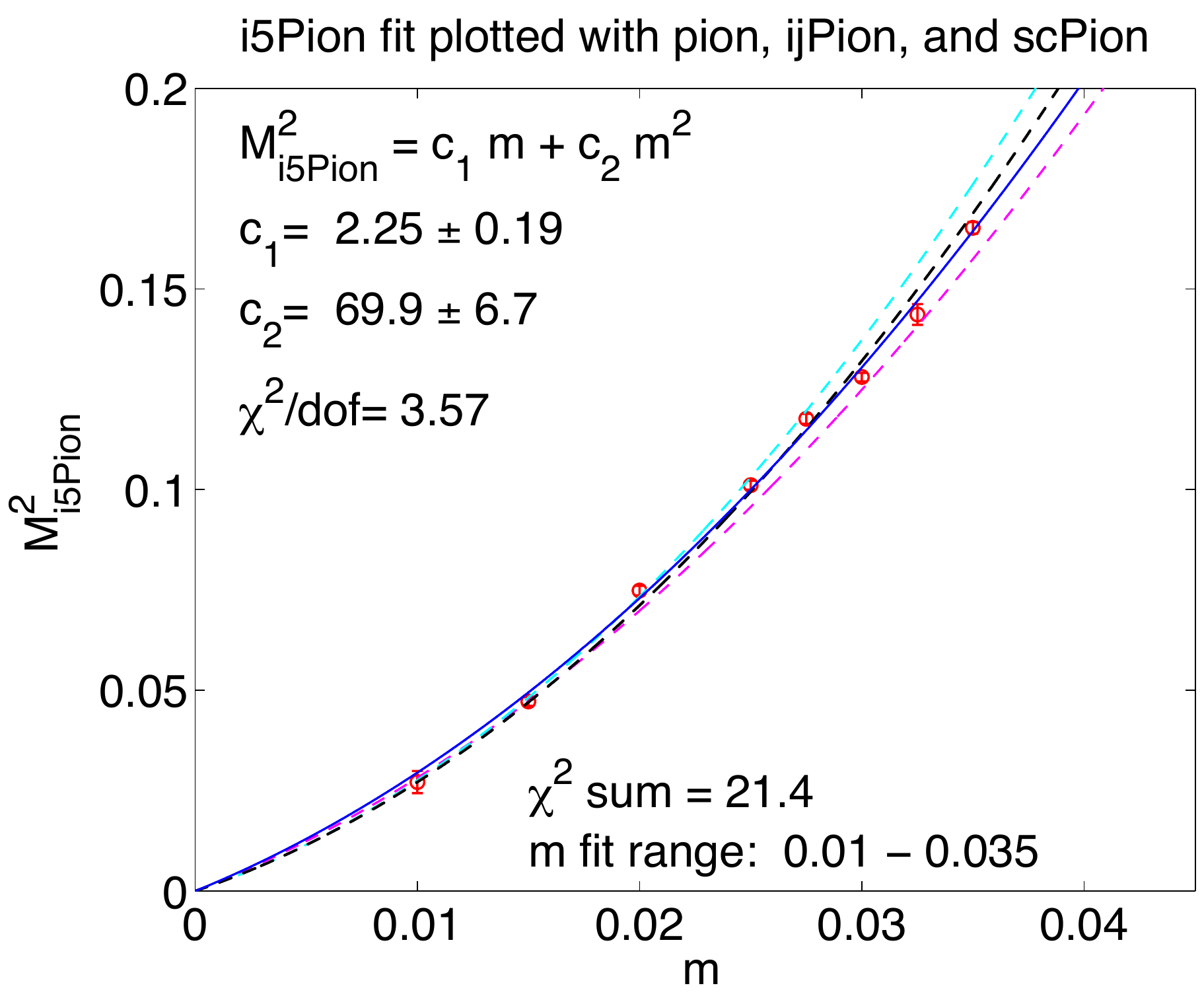}&
\includegraphics[height=4.6cm]{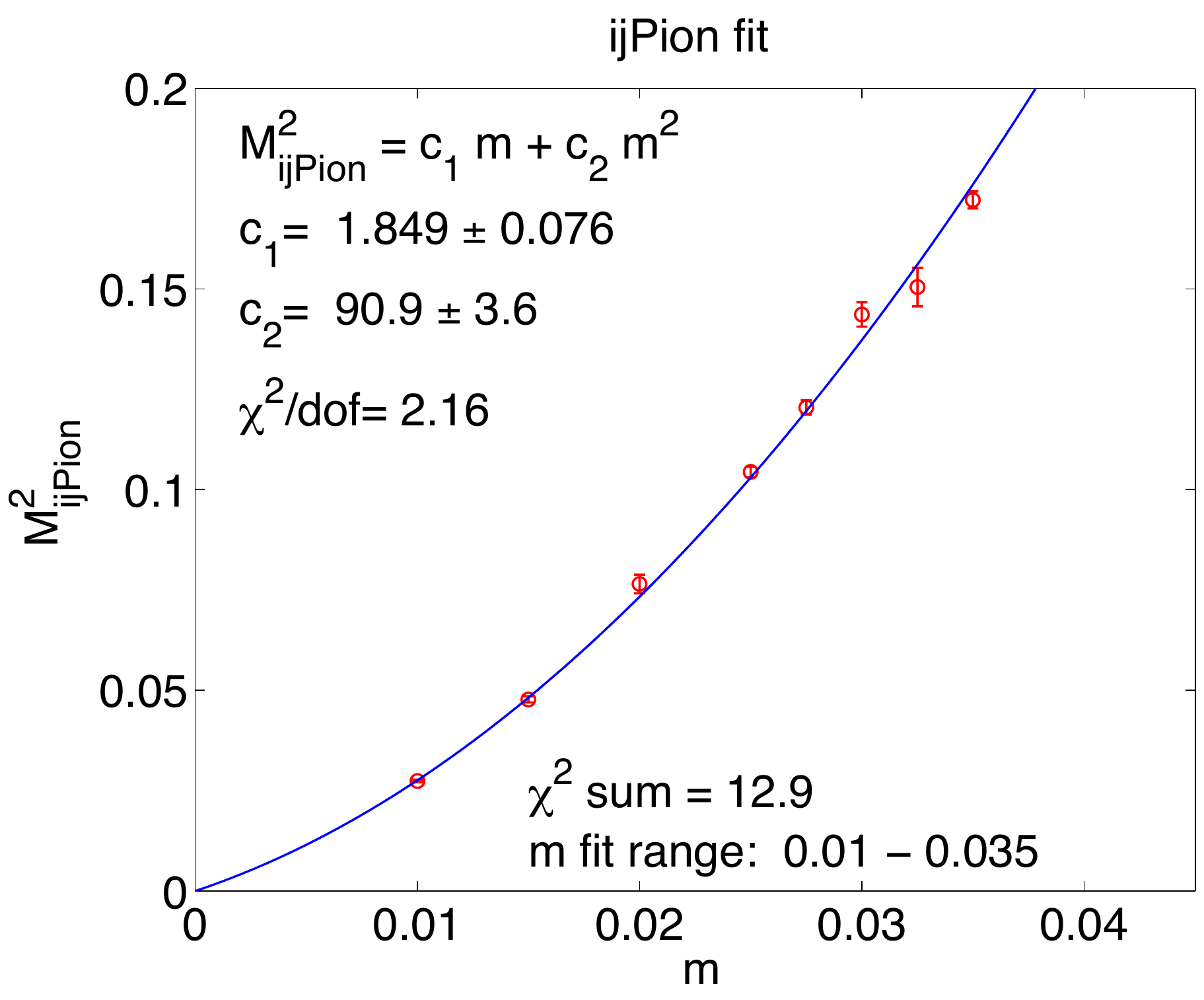}
\includegraphics[height=4.6cm]{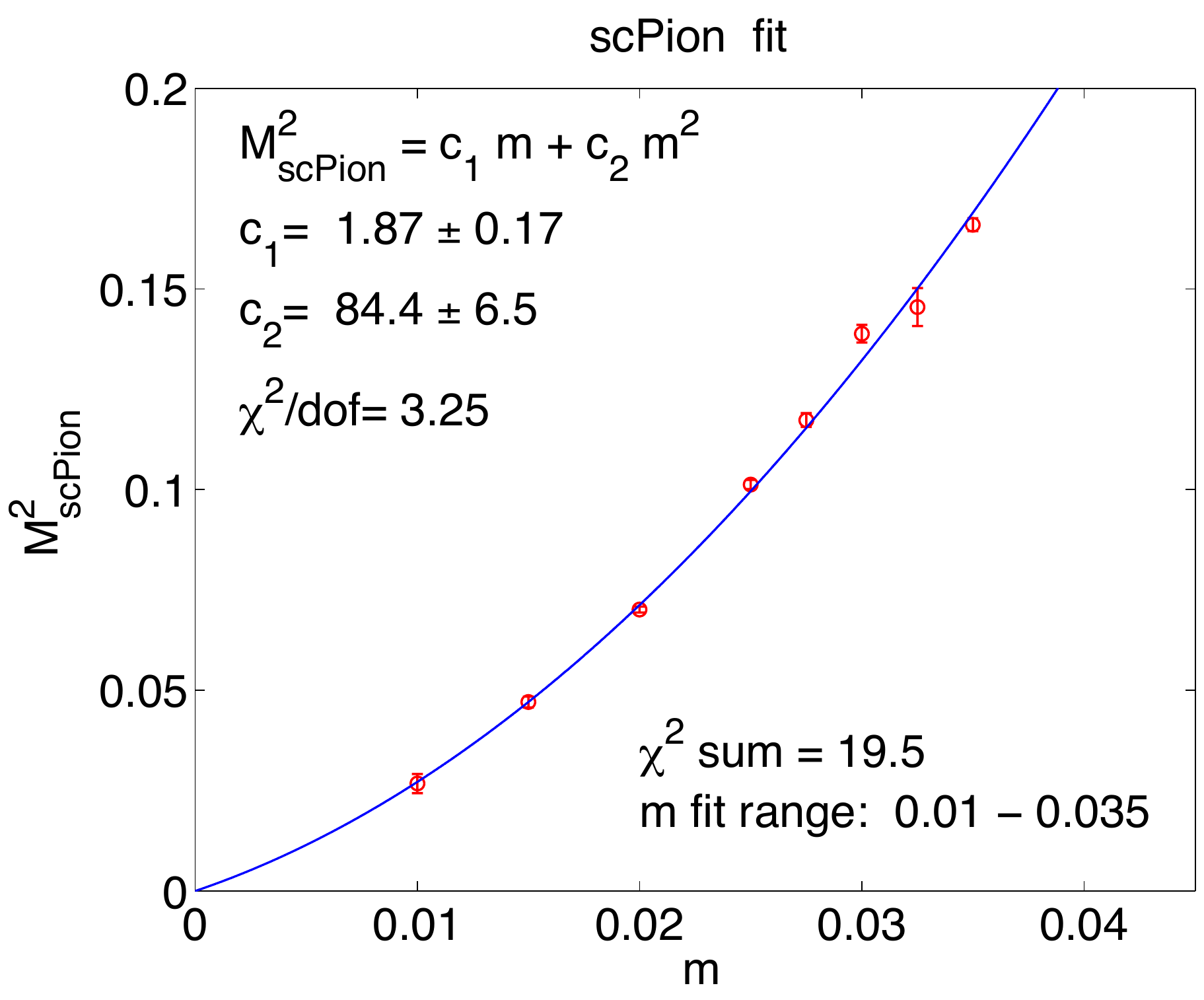}
\end{tabular}
\caption{{\footnotesize The non-Goldstone pion spectrum is shown. The composite left plot displays the i5Pion 
data and fit together with  fits to the Goldstone pion (magenta), i5Pion (solid blue), scPion (black), and ijPion (cyan).}}
\end{center}
\label{fig:Spectrum}
\vskip -0.2in
\end{figure*}

Figure 1 shows the Goldstone pion and $F_\pi$ as a function of the fermion mass $m$ in the range where we can
reach the infinite volume limit with  confidence.
The power functions of the fitting procedure in $m$ contain the analytic contributions of the fourth order 
chiral Lagrangian to $M_\pi$ and $F_\pi$.  Although we could fit the pion spectrum with the 
logarithmic term included, its significance remains unclear.
The rapid variation of $F_\pi$ with $m$ clearly shows that we would need 
a dense set of data in the $m=0.003-0.01$ range to reach chiral logs at this gauge coupling. 
This requires lattice volumes
well beyond the largest size $48^3\times 96$ which we could deploy in our simulations. 

Efforts were made for extrapolations to the infinite volume limit. 
%Finite volume scaling is very different under the hypotheses of the two different scenarios. 
%The power-like volume dependence in the conformal phase is driven by the mass anomalous dimension 
%while in the  ${\rm\chi SB}$ phase, as used here, the dominant exponential dependence on the volume is dominated 
%by the lightest pion mass.
At the lowest three $m$ values, for  finite volume corrections to  $M_\pi$ and $F_\pi$, and for 
all other states, we used the form
\begin{eqnarray}
 &&M_\pi(L_s,\eta) = M_\pi  \biggl [1+\frac{1}{2N_f}\frac{M^2}{16\pi^2F^2}\cdot\tilde g_1(\lambda,\eta) \biggr ] ,\\
 &&    F_\pi (L_s,\eta) = F_\pi \biggl [1-\frac{N_f}{2}\frac{M^2}{16\pi^2F^2} \cdot\tilde g_1(\lambda,\eta) \biggr ] ,      
\label{eq:MpiL}
\end{eqnarray}
where $\tilde g_1(\lambda,\eta)$ describes the finite volume corrections with
$\lambda=M_\pi\cdot L_s$ and aspect ratio $\eta=L_t/L_s$ from the lightest pion wrapping around the lattice
and coupled to the measured state~\cite{Leutwyler:1987ak}.  
The form of $\tilde
g_1(\lambda,\eta)$ is a complicated infinite sum which contains Bessel
functions and requires numerical evaluation. Since we are not in the chiral log regime, the prefactor of
the $\tilde g_1(\lambda,\eta)$ function was replaced by a fitted coefficient. The leading term of  the function
$\tilde g_1(\lambda,\eta)$ is a special exponential Bessel function $K_1(\lambda)$ which dominates in the simulation range.
The fitting procedure could be viewed as the approximate  leading treatment of  the pion which wraps around the finite volume,
whether in chiral perturbation theory, or in L\"uscher's non-perturbative finite volume analysis~\cite{Luscher:1985dn} 
which does not require the chiral limit
as long as the pion is the lightest state dominating the corrections.
The $M_\pi L_s > 4$ lore for volume independence is clearly not applicable in the
model. We need $M_\pi L_s > 8$ to reach volume independence.
The infinite volume limits of $M_\pi$ and $F_\pi$  for each $m$ were determined self-consistently from the fitting
procedure using Eqs.~(1,2) based on 
a set of $L_s$ values with  representative fit results shown in Figures 1 and 4. In the higher $m$ range finite volume effects
were hard to detect and even for the lowest $m$ values sometimes volume dependence was not detectable for the largest
lattice sizes. 

\begin{figure*}[hpt]
\begin{center}
\begin{tabular}{cc}
\includegraphics[height=6.8cm]{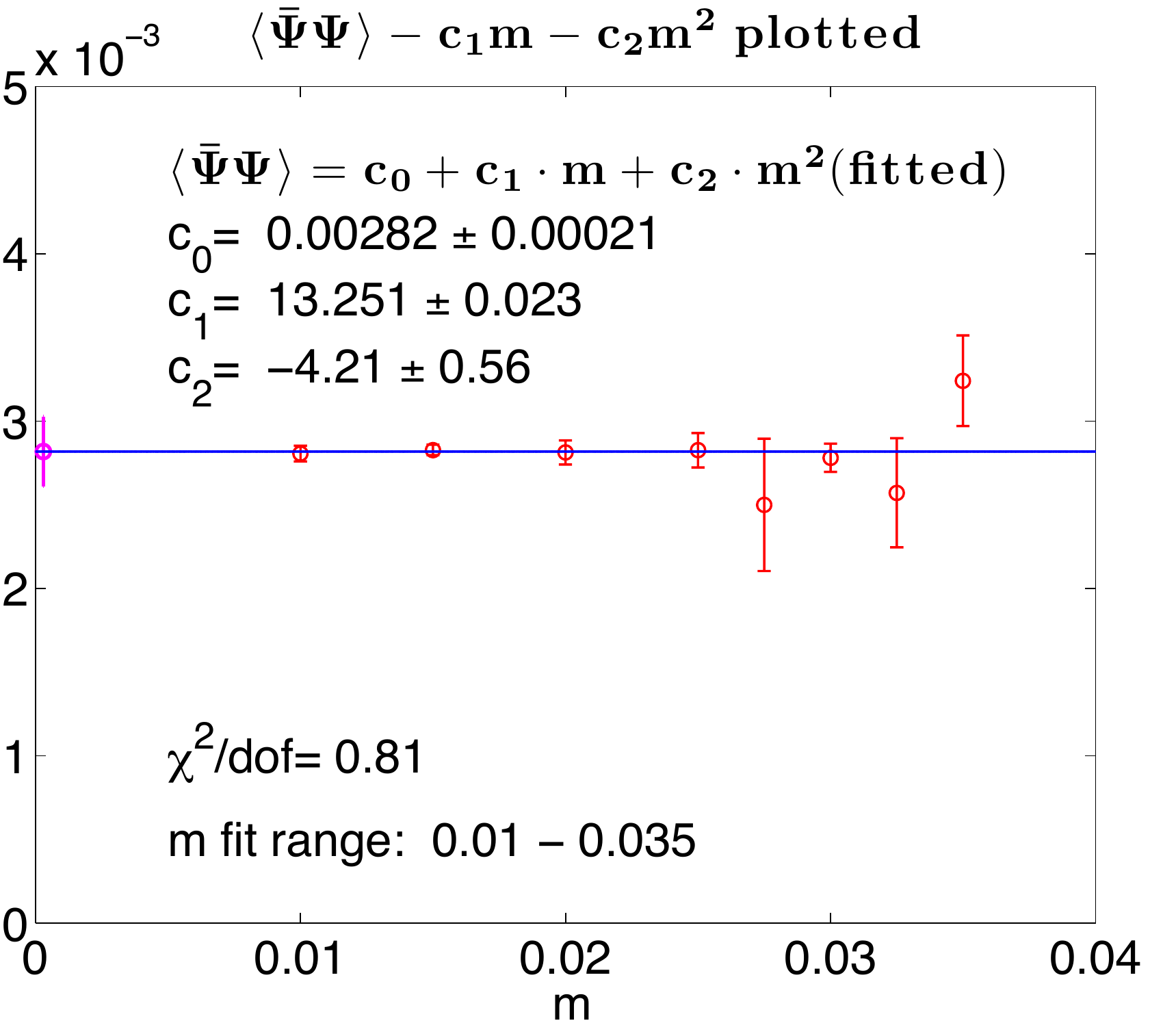}&
\includegraphics[height=6.8cm]{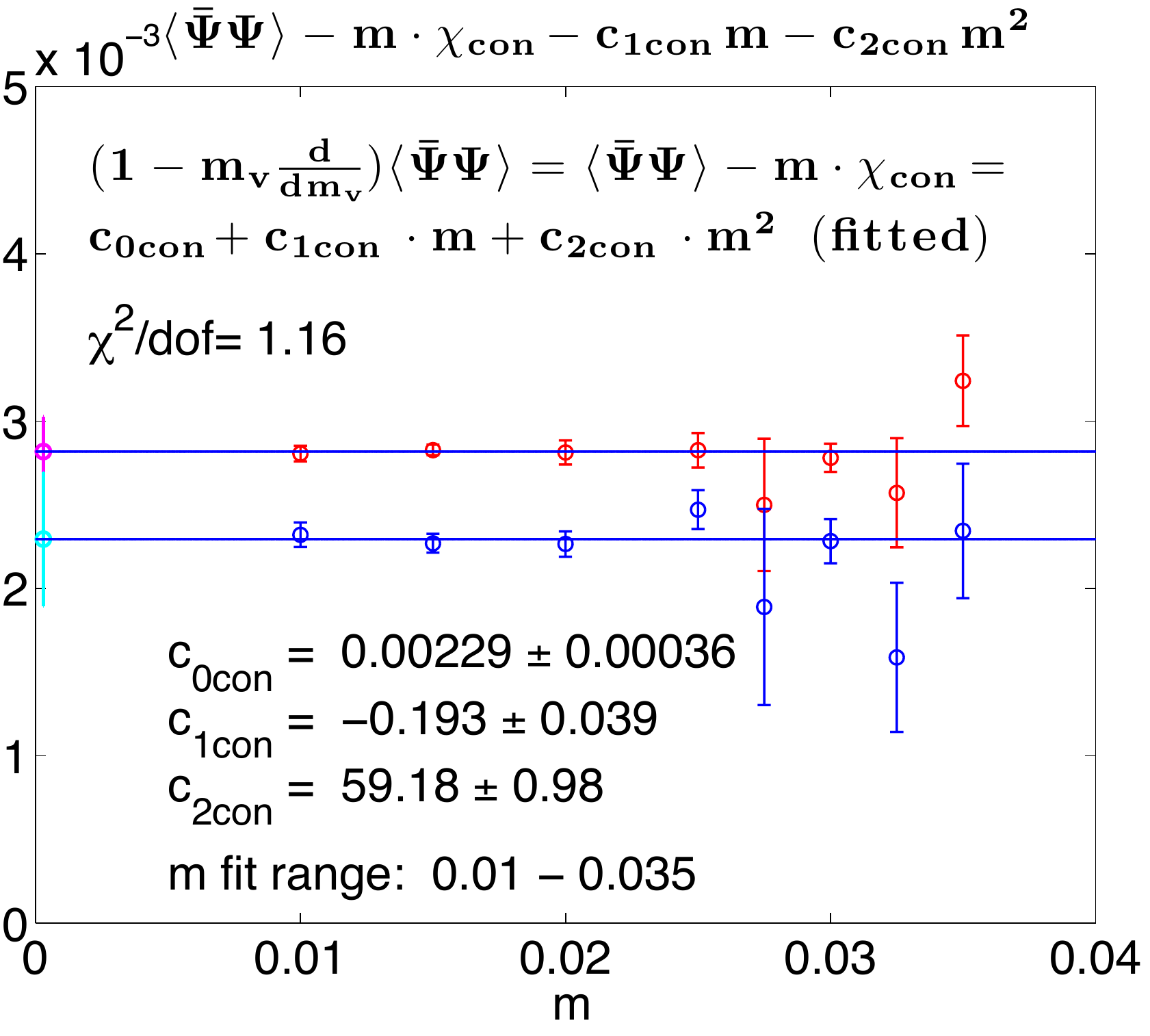}

\end{tabular}
\caption{{\footnotesize The chiral condensate is shown on the left. After the coefficients of the quadratic 
fitting function were determined, the plot shows data for $\langle \bar{\psi}\psi\rangle - c_1m+c_2m^2$ for better visual
display of the chiral limit at $m=0$ and its error (in magenta) coming from the fitted $c_0$ constant part. 
The right side is the quadratic fit to $\langle \bar{\psi}\psi\rangle - m\cdot\chi_{con}$ which is defined in the text
and directly measured from zero momentum sum rules and independently from functions of the inverse staggered fermion
matrix. The fitting function is $c_{0con} + c_{1con}m+ c_{2con} m^2$. 
 After the coefficients of the quadratic 
fitting function were determined, the plot shows data (blue points) for $\langle \bar{\psi}\psi\rangle - m\cdot\chi_{con} - c_{1con}m-c_{2con} m^2$ for better visual
display of the chiral limit at $m=0$ and its error (in cyan) coming from the fitted $c_{0con}$ constant part. 
For comparison, the left side plot is redisplayed showing consistency between the two different and independent
determinations of the chiral condensate in the chiral limit.
For any given $m$  always the largest volume chiral condensate data is used since 
the finite volume analysis is not complete.
We will continue extended systematics at the lowest two or three $m$ values which play an important role in the analysis.}}
\end{center}
\label{fig:PbPNf12}
\vskip -0.2in
\end{figure*}

Non-Goldstone pion spectra, quite different from those found in QCD, are shown in Figure 2 using standard notation.
They are not used in our global analysis.
The three states we designate as i5Pion,
ijPion and scPion do not show any noticeable taste breaking or residual mass in the $m\rightarrow 0$ chiral limit. 
The scPion is degenerate with the i5Pion and both are somewhat split from the true Goldstone pion. 
The ijPion state is  further split as expected but the overall taste breaking is very small across  the four pion states. 
This is a fairly strong indication that the coupling 
constant $\beta=2.2$ where all runs are performed is close to the continuum limit.
A very small residual mass at $m=0$ is not excluded for some non-Goldstone pion states depending on the
details of the fitting procedure. 

The staggered meson and baryon states and correlators we use are defined in ~\cite{Ishizuka:1993mt}.
For example, what we call the scPion and the $f_0$ meson are
identified in correlator I of Table 1 in  ~\cite{Ishizuka:1993mt}. Similarly,  the i5Pion is from correlator VII, the ijPion is from correlator VIII, the rho and A1 mesons are
from correlator III of Table 1. We measure the Goldstone pion in two different ways, with one of them defined  in correlator II of Table 1 in the reference. The nucleon state and its 
parity partner are defined in correlator I of Table 2 in ~\cite{Ishizuka:1993mt}.

\subsection{Chiral condensate}

The chiral condensate $\langle \bar{\psi}\psi\rangle$ summed over all flavors
has the spectral representation~\cite{Banks:1979yr}
\bea
&&\langle \bar{\psi}\psi\rangle = -2m\cdot\int^{\mu}_0\frac{d\lambda\rho(\lambda)}{m^2+\lambda^2}\nn\\
&&\hskip 0.3in=-2m^5\cdot\int^{\infty}_\mu\frac{d\lambda}{\lambda^4}\frac{\rho(\lambda)}{m^2+\lambda^2}
+c_1\cdot m + c_3\cdot m^3 
\eea
where the UV-divergent integral is written in a twice-subtracted form in the second line~\cite{Leutwyler:1992yt}.
The UV contribution, which is divergent when the cutoff  $a^{-1}$ is removed, has a linear term  $\approx a^{-2}\cdot m$  and 
there is a third-order term  $\approx\! m^3$ which is hard to detect for small $m$  and survives even in the free theory limit.
The IR finite contributions to the chiral Lagrangian
have a constant term $\approx\! B F^2$, a linear term $\approx\! B^2\cdot m$, a quadratic term $\approx B^3F^{-2}\cdot m^2$, and higher 
order terms in addition to
logarithmic corrections generated from chiral loops~\cite{Bijnens:2009qm}.

We kept a constant IR term, the linear term with UV and IR contributions, and the quadratic IR term in our fitting procedure of  $\langle \bar{\psi}\psi\rangle$.
The quadratic fit in Figure 3
gives a small non-vanishing condensate in the chiral limit which is roughly consistent with 
the GMOR~\cite{GellMann:1968rz} relation  $\langle\bar{\psi}\psi\rangle=12F^2B$ 
with the measured low value of $F$  and  O(1) value for $B$ which correspond to the Goldstone pion fits
in Figure 1. 
The deficit between the two sides of the GMOR relation is sensitive to the fitting procedure and  the uncertain determination of  $B$.
The quadratic term in the fit is a relatively small contribution and trying to identify chiral logs is 
beyond the scope of our simulation range. 

For an independent determination, we also studied the subtracted chiral condensate operator defined with the help
of the connected part $\chi_{con}$ of the chiral susceptibility $\chi$,
\bea
&&(1-m_v\frac{ d}{dm_v}) \langle\bar\psi\psi\rangle\mid_{m_v=m} = \langle\bar\psi\psi\rangle - m\cdot\chi_{con}~, \\
&&\chi =\frac{ d}{dm} \langle\bar\psi\psi\rangle = \chi_{con} + \chi_{disc}~, \nn \\
&&  \chi_{con}=\frac{ d}{dm_v}\langle\bar\psi\psi\rangle_{pq}\mid_{m_v=m}~ .\nn
\eea
The derivatives  $d/dm$ and  $d/dm_v$ are taken at fixed gauge coupling $\beta$. The derivative
$d/dm_v$ is defined in the partially quenched functional integral of $\langle \bar{\psi}\psi\rangle_{pq}$
with respect to the valence mass $m_v$
and the limit $m_v=m$ is taken after differentiation.
The removal of the derivative term significantly reduces the 
dominant linear part of the $\langle \bar{\psi}\psi\rangle$ condensate. 
We find it reassuring that the two independent determinations give consistent non-vanishing results in the chiral limit
as clearly shown in Figure 3.

\begin{table}[ht]
\centering
\caption{The chiral condensate $\langle \bar{\psi}\psi\rangle$ and $\langle \bar{\psi}\psi\rangle - m\cdot\chi_{con}$, defined in the text
and directly measured from zero momentum sum rules and independently from functions of the inverse staggered fermion
matrix, are tabulated and used in the fits of Figure 3.}
\vskip 0.05in
\resizebox{0.4\textwidth}{!}{
\begin{tabular}{|l|l|l|l|l|}
\hline\hline
mass & lattice &$\hskip 0.25in\langle \bar{\psi}\psi\rangle$&$\langle \bar{\psi}\psi\rangle - m\cdot\chi_{con}$\\
\hline
\hline  0.0100& $48^3\times  96$ & 0.134896(47) &  0.006305(73) \\ 
\hline  0.0150& $48^3\times  96$ & 0.200647(31) &  0.012685(56) \\ 
\hline  0.0200& $40^3\times  80$ & 0.266151(72) &  0.022069(76) \\ 
\hline  0.0250& $32^3\times  64$ & 0.33147(10)   &  0.03462(12)   \\ 
\hline  0.0275& $24^3\times  48$ & 0.36372(40)   & 0.04133(59)    \\ 
\hline  0.0300& $32^3\times  32$ & 0.396526(84) &  0.04974(13)   \\ 
\hline  0.0325& $24^3\times  48$ & 0.42879(33)   &  0.05781(45)   \\ 
\hline  0.0350& $24^3\times  48$ & 0.46187(27)   &  0.06807(40)   \\ 
\hline 
\end{tabular} }
\label{table:2}
\end{table}

\begin{figure*}[hpt]
\begin{center}
\begin{tabular}{ccc}
\includegraphics[height=4.7cm]{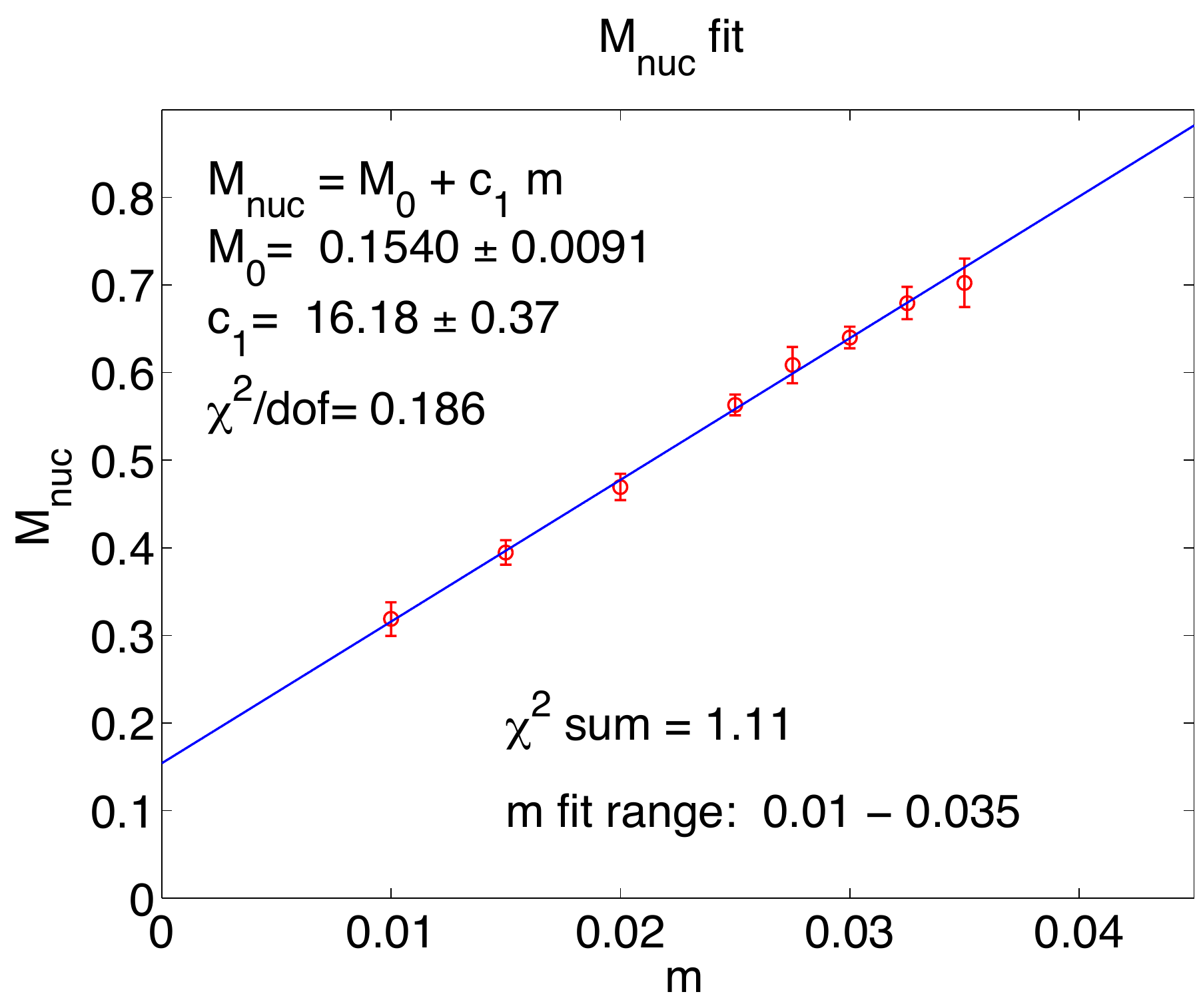}&
\includegraphics[height=4.7cm]{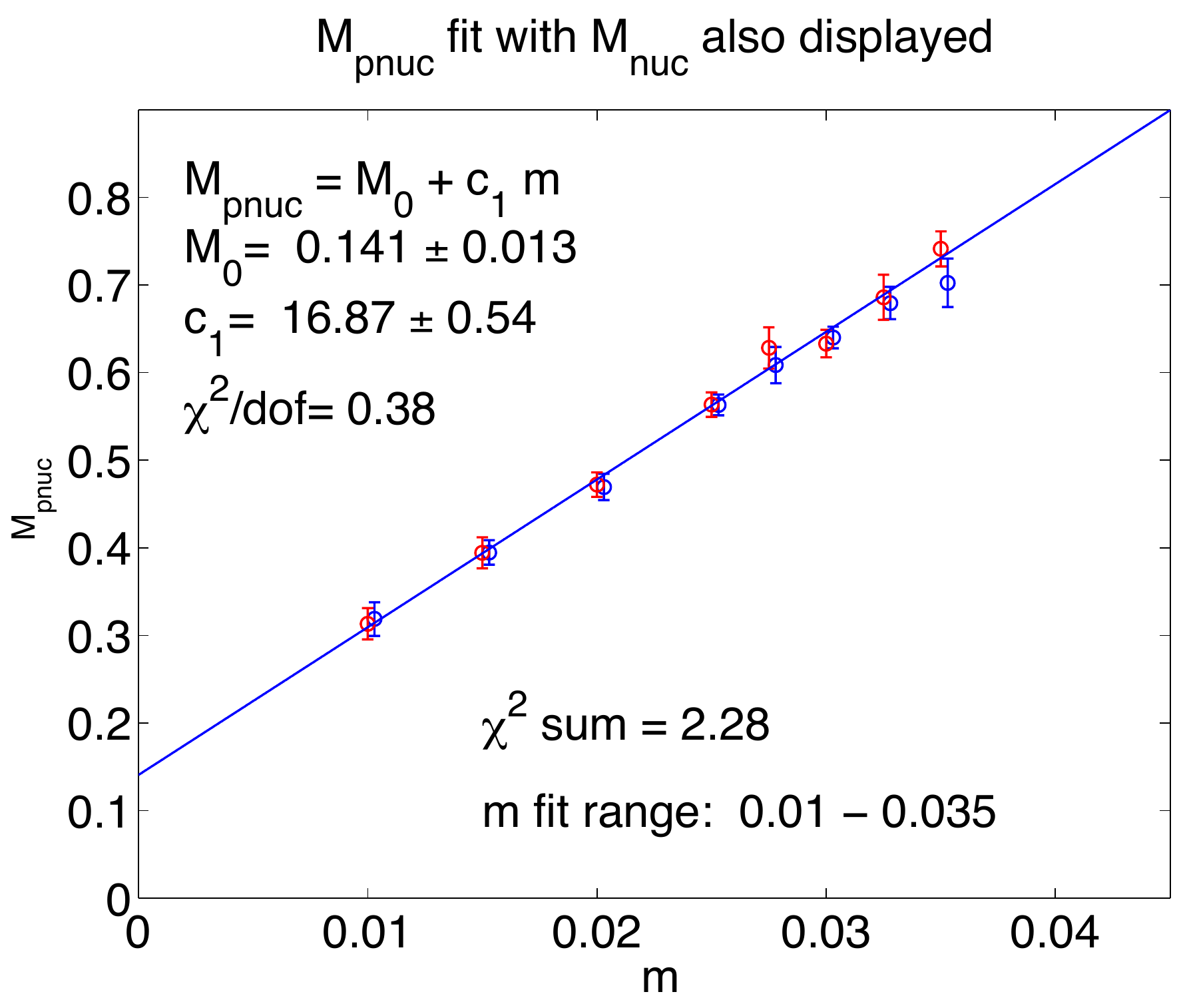}&
\includegraphics[height=4.7cm]{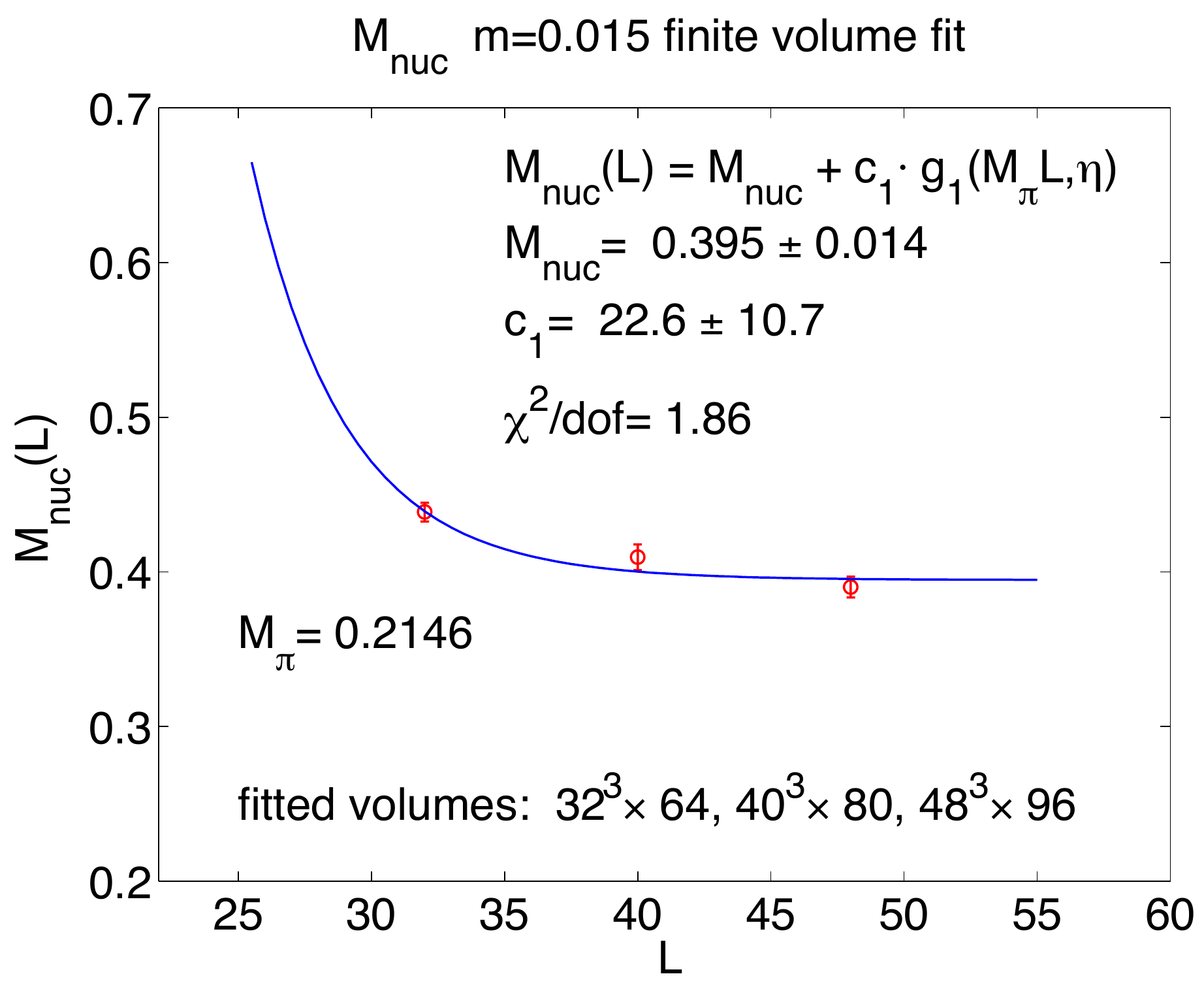}
\end{tabular}
\caption{{\footnotesize Nucleon and its parity partner are fitted to the constant plus linear form which are  the leading contributions of the chiral 
Lagrangian. The blue points in the middle plot are the
replotted nucleon data from the left to show the degeneracy of the two states. The plot on the right shows a representative finite volume fit.}}
\end{center}
\label{fig:Nucleon}
\vskip -0.2in
\end{figure*}

It should be noted that the $M_\pi$ values in the fitting range of $m$  in our analysis are {\em below}
the fitting range of  previous $N_f=12$ work on the chiral condensate work 
with considerably more uncertainty  from using the higher range~\cite{Deuzeman:2009mh}.  
In all fits we were on a fine-grained lattice in the pion mass range $aM_{\pi}=0.16-0.39$ 
and rho mass range $M_{\rho}=0.2-0.47$. In contrast, the previous study~\cite{Deuzeman:2009mh}
which reported conformal behavior was in the $aM_{\pi}=0.35-0.67$ range and rho mass range $M_{\rho}=0.39-0.77$.
Although our new results should be made even more definitive with higher accuracy and  better control on  
the systematics, the evidence is quite suggestive
for a small non-vanishing chiral condensate in the chiral limit.

\subsection{Composite hadron spectrum in the chiral limit}

\begin{figure}[htb]
\begin{center}
\begin{tabular}{cc}
\includegraphics[height=3.5cm]{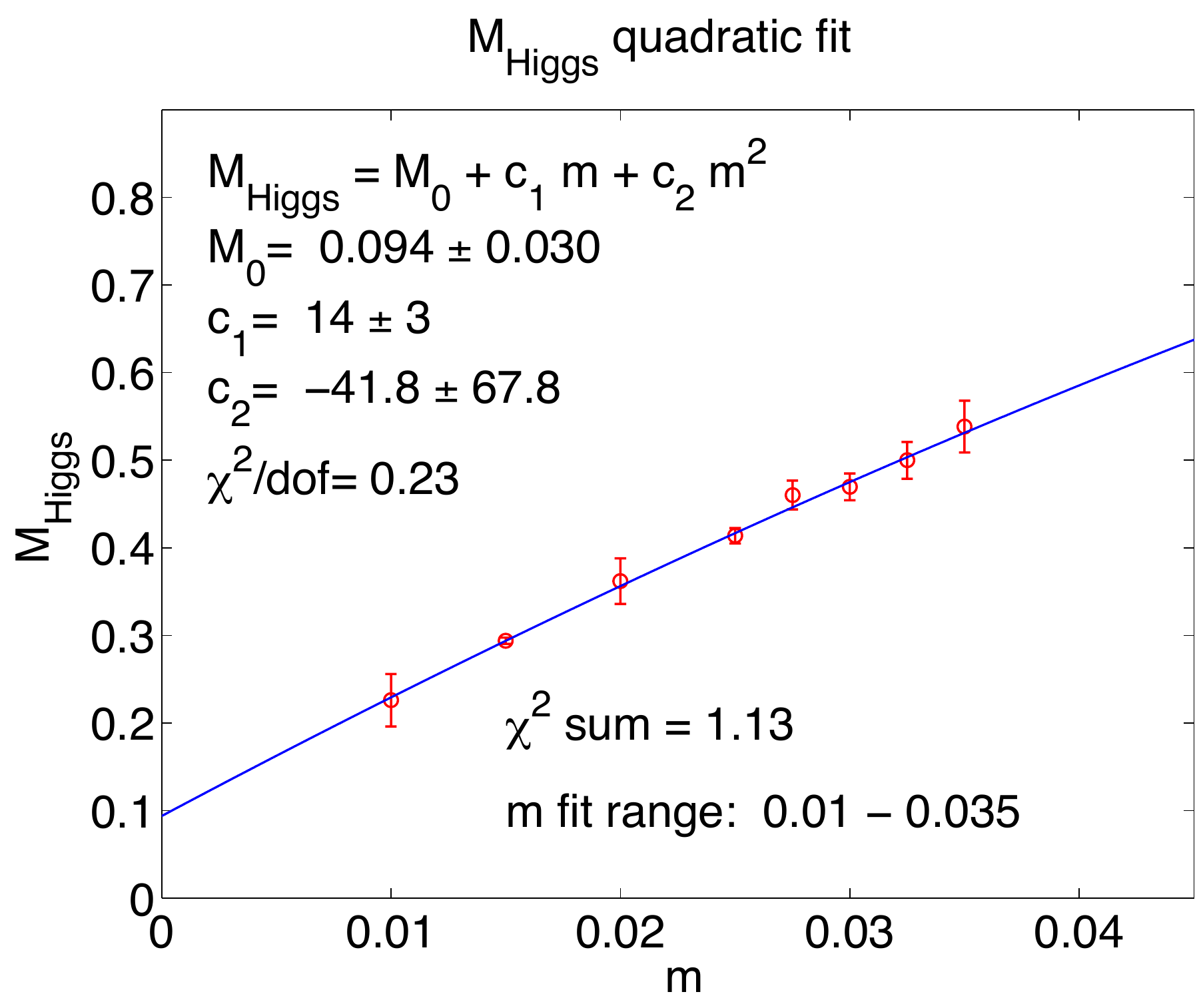}&
\includegraphics[height=3.5cm]{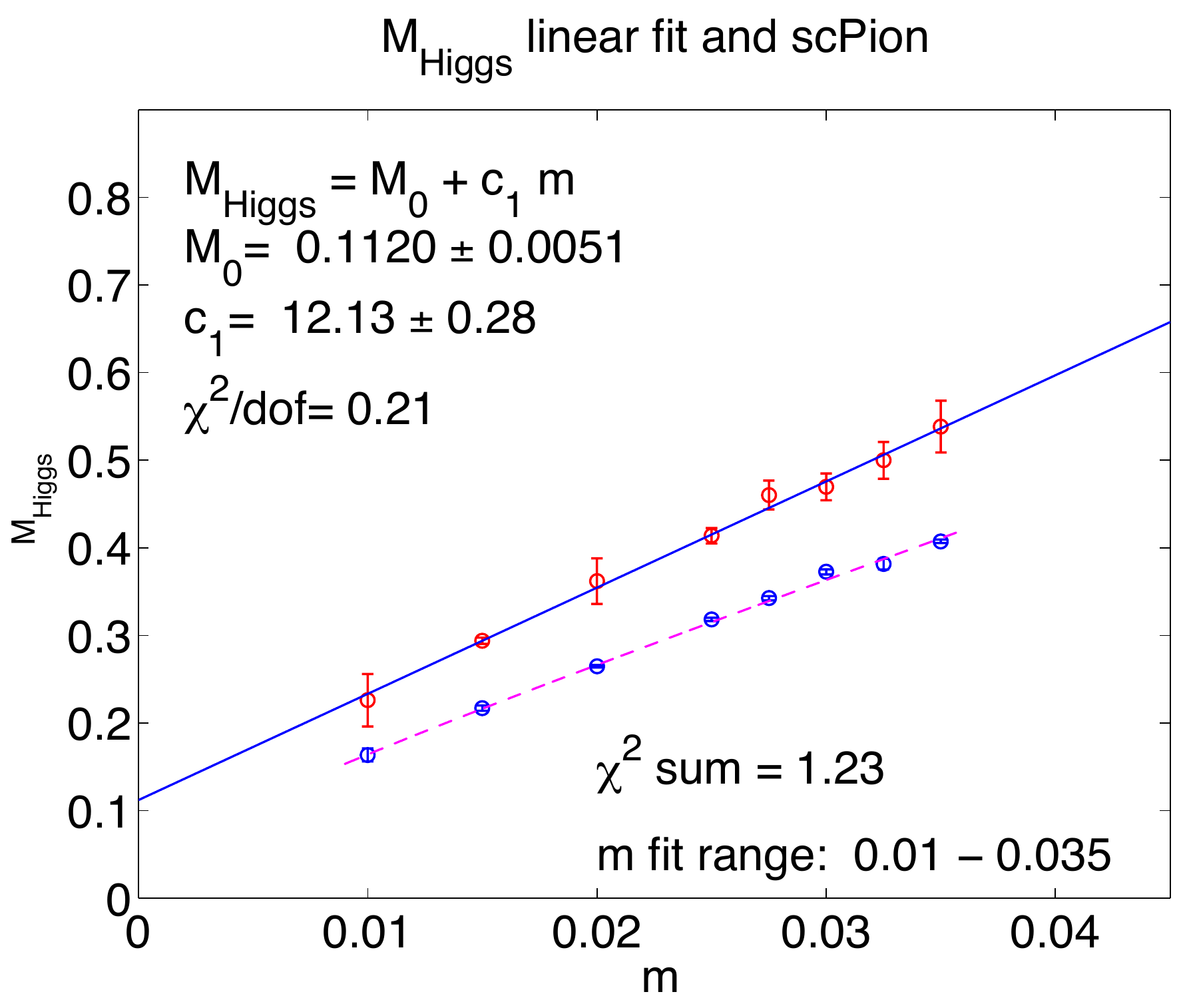}
\end{tabular}
\caption{{\footnotesize The Higgs ($f_0$) state  and its splitting from the scPion state are shown. The linear fit on the right
works well for the Higgs ($f_0$)
state with little change when a quadratic term is included on the left. The blue scPion
data points on the right and the dashed magenta fit show the fit to the scPion state.  
The Higgs will become a resonance in the chiral limit, the missing disconnected part also contributing, so that Higgs
predictions will be challenging in future work.}}
\end{center}
\label{fig:Higgs}
\vskip -0.2in
\end{figure}
\begin{figure}[htpb]
\begin{center}
\begin{tabular}{cc}
\includegraphics[height=3.5cm]{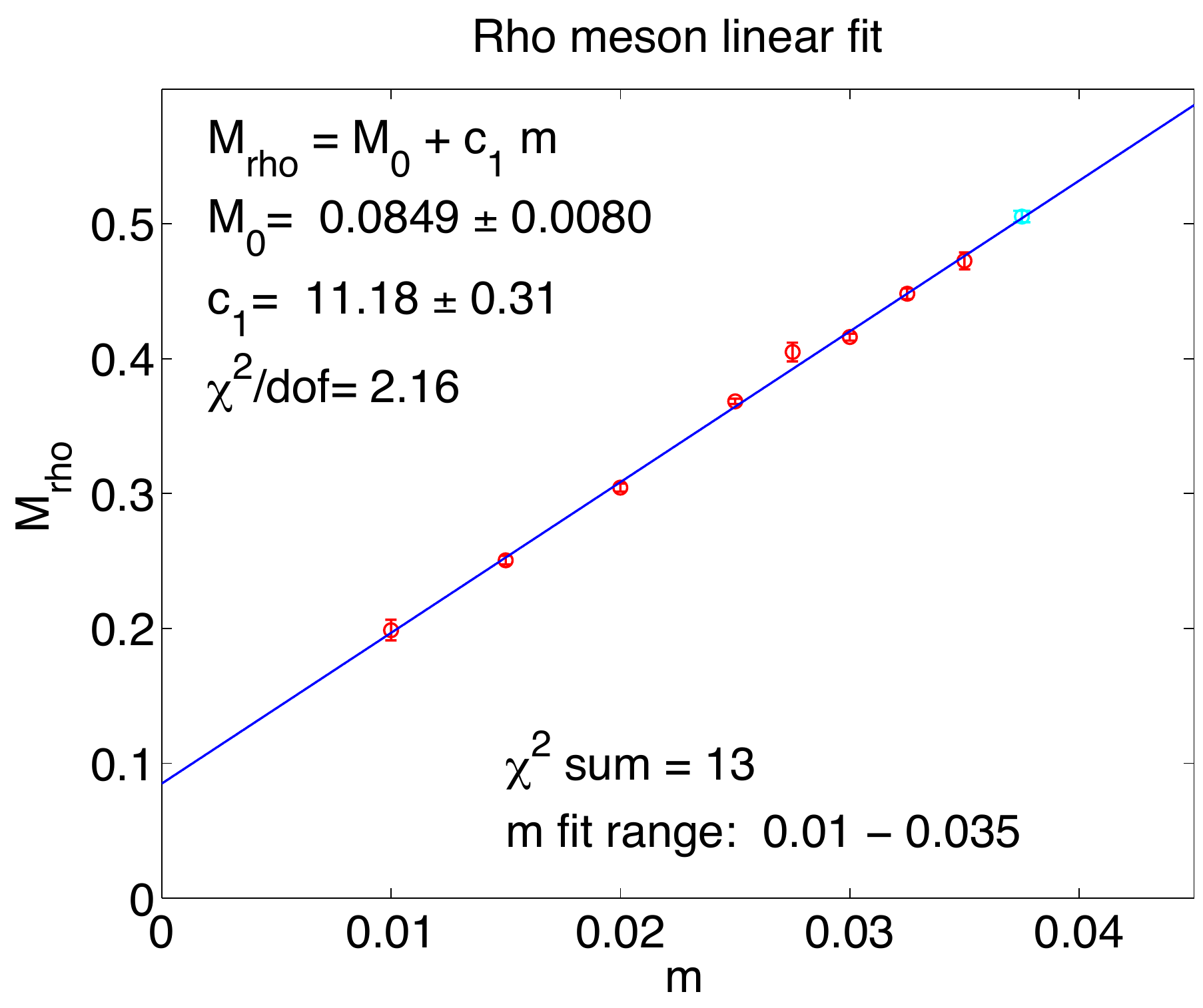}&
\includegraphics[height=3.5cm]{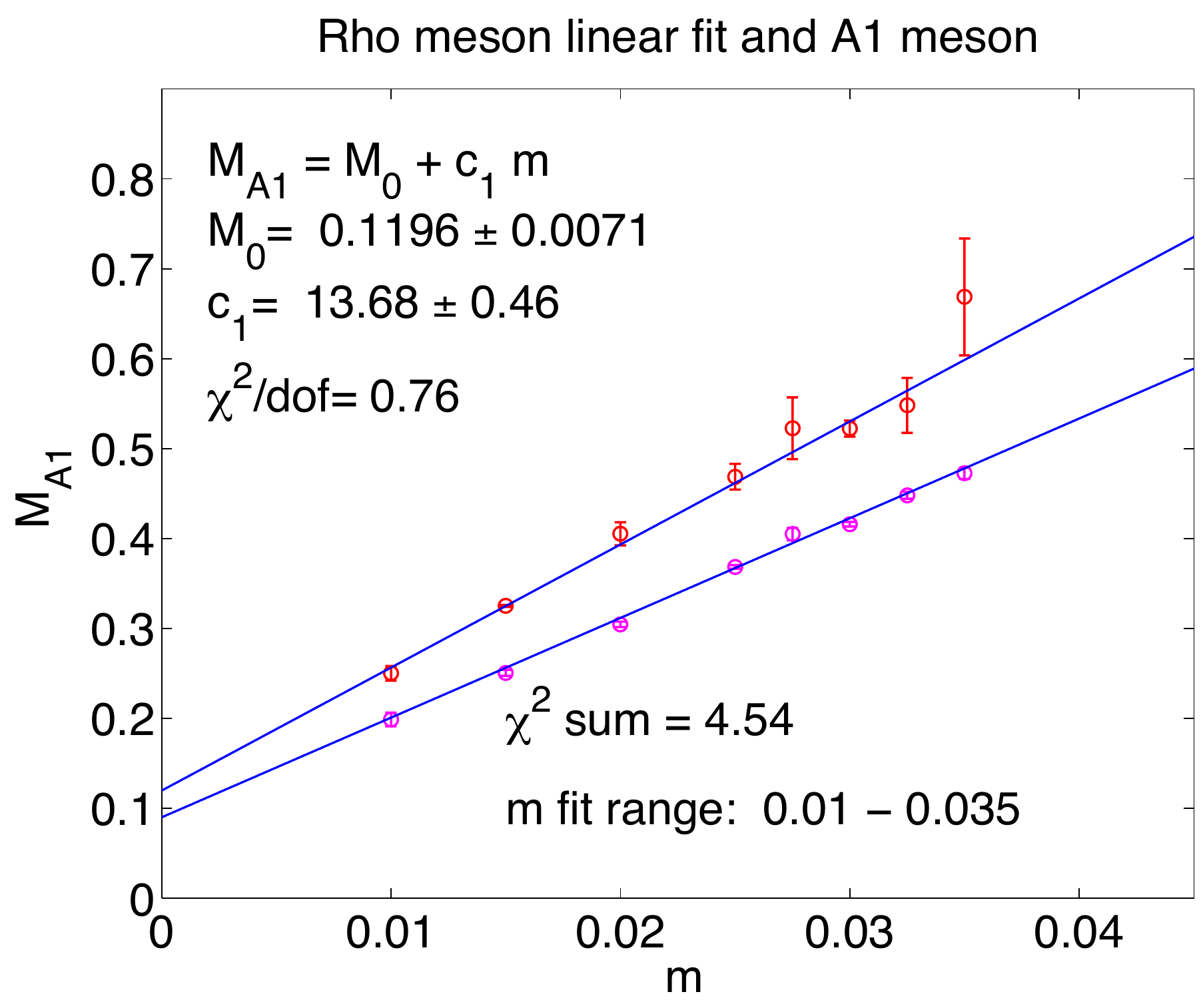}\\
\end{tabular}
\caption{{\footnotesize Rho meson and its splitting from the A1 meson are shown. 
On the right side the magenta points reproduce the data of the rho meson from the left together with its linear fit. The fit
parameters on the right show the linear fit to the $A_1$ meson.}}
\end{center}
\label{fig:Rho}
\vskip -0.1in
\end{figure}

\begin{table*}[!htpb]
\caption{V(R) tabulated at fermion masses $m=0.010$ and $m=0.015$ for lattice volume $48^3\times 96$, and at $m=0.020$ for lattice volume $40^3\times 80$.}
\resizebox{\textwidth}{!}{
\centering
\begin{tabular}{|c|c|c|c|c|c|c|c|c|c|c|c|}
\hline\hline
m\textbackslash R & 4  & 5 & 6 & 7 & 8 & 9 & 10 & 11 & 12 & 13 & 14 \\
\hline \hline 
0.010  & 0.20005(49)  & 0.22686(84) & 0.24638(12)  &
0.26000(28) & 0.27059(55) & 0.27957(82) & 0.2872(10) & 0.2933(21) &
0.2979(42) & 0.30771(31) & 0.31250(82) \\
\hline 
0.015 & 0.20439(21) & 0.23332(35) & 0.25270(39) & 0.26737(85) &
0.2789(17) & 0.2892(28) & 0.30214(36) & 0.3129(11) & 0.3220(31) &
0.3289(12) & 0.33576(43) \\
\hline  
0.020 & 0.20819(39) & 0.2372(16) & 0.25961(99) & 0.27727(55) &
0.29132(74) & 0.3040(11) & 0.31718(24) & 0.32862(31) & 0.33973(78) &
0.34921(77) & 0.3543(55) \\
\hline 
m\textbackslash R &  15 & 16 & 17 & 18 & 19 & 20 & 21 & 22 & 23 & 24 & \\
\hline 
0.010 & 0.31755(43) & 0.32186(78) & 0.3263(19) & 0.3308(23) &
0.3339(40) & 0.3364(47) & 0.3417(27) & 0.3453(29) & 0.3466(62) &
0.3554(25) & \\
\hline 
0.015 & 0.34295(46) & 0.35050(37) & 0.35863(78) & 0.36506(45)
& 0.36928(69) & 0.3708(31) & 0.3741(55) & 0.3817(59) & 0.3897(71) & & \\
\hline
0.020 & 0.3625(58) & 0.3768(24) & 0.3939(107) & 0.3946(10) & 0.4026(13)
& 0.4085(30) & & & & & \\ 
\hline 
\end{tabular} }
\label{table:3}
\end{table*}

\begin{figure*}[!htpb]
\begin{center}
\begin{tabular}{ccc}
\includegraphics[width=0.32\textwidth]{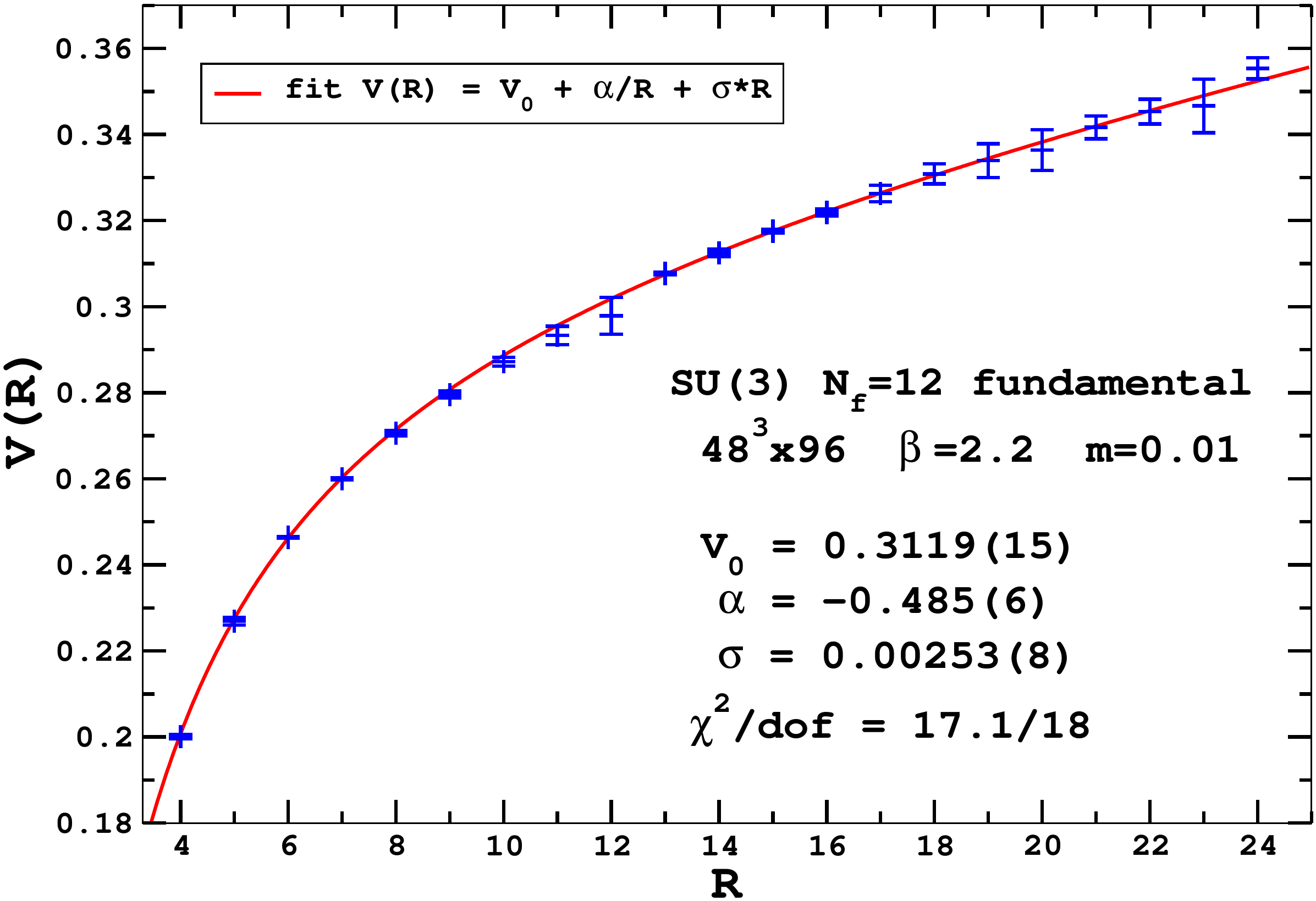}&
%\hspace{3.5cm}
\includegraphics[width=0.32\textwidth]{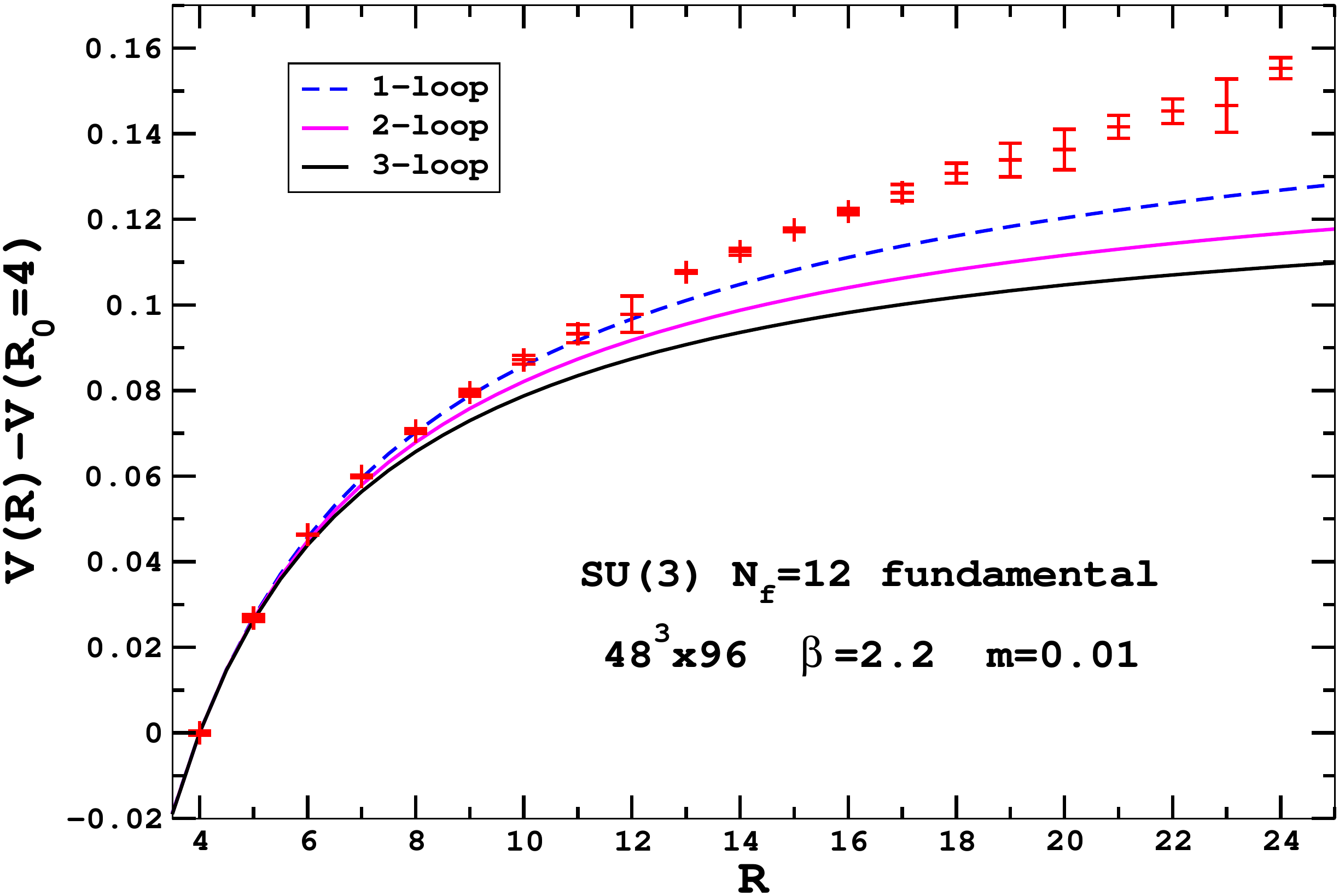} &
%\hspace{3.5cm}
\includegraphics[width=0.32\textwidth]{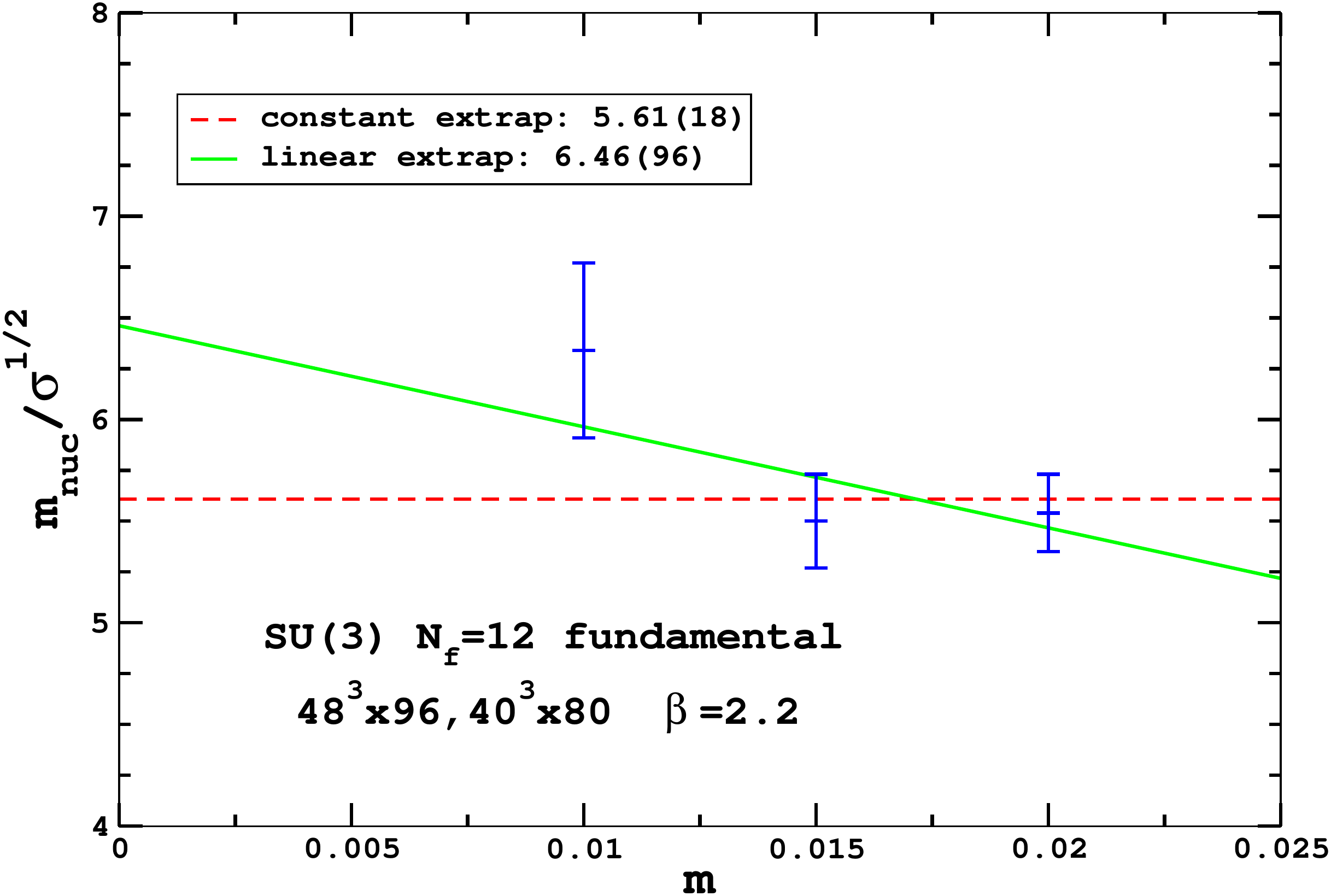} 
\end{tabular}
\caption{{\footnotesize $V(R)$ data and fit for  $m=0.01$ is plotted on the left and comparison with
perturbation theory is shown in the middle plot. The right side plot shows
the string tension  measured in nucleon mass units at $m=0.01,0.015,0.02$ and extrapolated to the chiral limit. 
The finite nucleon mass  gap in the chiral limit implies a finite  string tension at $m=0$. }}
\label{fig2}
\end{center}
\end{figure*}

It is important to investigate the chiral limit of other composite hadron states. They further test the 
mass splittings between physical states as the fermion mass $m$ is varied and the measured hadron masses are subjected to chiral analysis
in the $m\rightarrow 0$ limit for important residual mass gaps above the vacuum after infinite volume extrapolation. 
Hadron masses also provide useful information on parity splits in several channels.
One composite state of great interest is the Higgs particle, if there is a chiral condensate close to 
the conformal window. We will briefly review new results on  the nucleon state with its parity partner, the isospin partner
of the Higgs ($f_0$) state, and the $\rho-A_1$ splitting. 

The fermion mass dependence of the nucleon and its parity partner is shown in Figure 4 
with finite volume analysis at one selected fermion mass $m=0.015$. 
The same finite volume fit is applied as 
described earlier for the pion state. The leading chiral linear term in the fermion mass $m$ extrapolates to a non-vanishing 
chiral limit. The parity partner is practically degenerate but this is not a surprise. Already with four flavors 
a near degeneracy was reported before by the Columbia group~\cite{Mawhinney:1996jk}. 

Figure 5 shows the fermion mass dependence of the Higgs particle without including the disconnected
part of the relevant correlator. Strictly speaking, therefore, the state is the $f_0$ meson with non-zero isospin.
Disconnected contributions in the correlator might shift the Higgs mass, an important issue left for future clarifications. 
Both the linear and the quadratic fits are shown together with the non-Goldstone
scPion which is split down from the Higgs ($f_0$) state. The two states would be degenerate in the chiral limit with unbroken symmetry.
The Higgs ($f_0$)  state extrapolates to a nonvanishing mass in the chiral limit with  an $M_{H (f_0)}/F$ ratio
between 10 and 15. 

Finally, Figure 6 shows the $\rho$-meson and its $A_1$ parity partner. Both states extrapolate to non-vanishing mass
in the chiral limit. The split remains significant for all fermion masses and in the chiral limit.

%\pagebreak
\subsection{String tension and running coupling from the static force}

There are several ways to define a renormalized gauge coupling, for
example, the Schr\"{o}dinger Functional scheme or from square Wilson
loops. We take the renormalized coupling as defined via the
quark-antiquark potential $V(R)$, extracted from $R \times T$ Wilson
loops where the time extent $T$ can be large. From the potential, one
defines the force $F(R)$ and coupling $\alpha_{qq}(R)$ as
\be
F(R) = \frac{dV}{dR} = C_F \frac{\alpha_{qq}(R)}{R^2}, \hspace{0.4cm}
\alpha_{qq}(R) = \frac{g^2_{qq}(R)}{4 \pi}.
\label{eq:F}
\ee
The coupling is defined at the scale $R$ of the quark-antiquark
separation, in the infinite-volume limit $L \rightarrow \infty$. This
is different from the scheme using square Wilson loops, where one has
$\alpha_W(R,L)$ and one can choose finite $R$ with $L \rightarrow \infty$,
or finite $L$ and fixed $R/L$ ratio. In the former case, these schemes
are related via 
\be
\alpha_{qq}(R) = \alpha_W(R) [ 1 + 0.31551 \alpha_W(R) + {\cal
  O}(\alpha_W(R)^2) ].
\label{eq:qqw}
\ee
The $\beta$ function in the qq scheme is known to 3-loops.
For SU(3) gauge theory with $N_f=12$ fundamental
flavors, the location of the infrared fixed point to 3-loop
order is $\alpha_{qq}^* = 0.3714...$ This is about 50\% of the
scheme-independent 2-loop value of $\alpha^*$, indicating that
higher order corrections beyond 3-loop might not be negligible.

 A range of lattice spacings, volumes and quark masses are studied in the running coupling project, we
show results for the largest volume $48^3 \times 96$ at $\beta=2.2$
and quark masses $m=0.01$ and 0.015 and for the $40^3 \times 80$ run at $m=0.02$.
To improve the measurement of $V(R)$, we use different levels
of APE-smearing to produce a correlation matrix of Wilson loops, the
lowest energy is extracted using the generalized eigenvalue method. We
also improve the lattice force, which is naively discretized as
$F(R+1/2) = V(R+1) - V(R)$. For the Symanzik gauge action, the
improvement is a relatively small effect, for example the naive value
$R+1/2 = 4.5$ is shifted to $4.457866...$

In Figure 7 on the left we show the measured $V(R)$ fitted to 
the form
\be
V(R) = V_0 + \frac{\alpha}{R} + \sigma R.
\ee
for $m=0.01$. The $m=0.015$ and $m=0.02$ runs are shown on the right of Figure 7. 
For all three masses, the resulting fits are good, with a clear signal of
linear dependence and an effective string tension $\sigma$. The string
tension decreases with the quark mass, its behavior in conjunction
with the mass spectrum in the chiral limit is under investigation and the first result is shown in the figure.
The finite nucleon mass  gap in the chiral limit implies a finite  string tension at $m=0$.

\begin{table}[ht]
\caption{$m_{\rm nuc}/\sqrt{\sigma}$.}
\centering
\begin{tabular}{|c|c|c|c|}
\hline\hline
$m$ & $\sigma$ & $\chi^2$/dof & $m_{\rm nuc}/\sqrt{\sigma}$ \\
\hline
0.01 & 0.002530(81) & 17.1/18 & 6.34(43) \\
0.015 & 0.005147(109) & 43.6/17 & 5.50(23) \\
0.02 & 0.007189(77) & 6.8/14 & 5.54(19) \\
\hline
\end{tabular} 
\label{table:4}
\end{table}

The renormalized coupling $\alpha_{qq}(R)$ is a derivative of the
potential $V(R)$ and hence more difficult to numerically measure via
simulations. The most accurate comparison between lattice simulations
and perturbation theory is directly of the potential $V(R)$
itself. This is naturally given by finite potential differences
\be
V(R) - V(R_0) = C_F \int_{R_0}^R \frac{\alpha_{qq}(R')}{R'^2} dR',
\ee
where $R_0$ is some reference point where $\alpha_{qq}(R_0)$ is
accurately measured in simulations. From this starting point, the
renormalized coupling runs according to perturbation theory, at some
loop order. The result is shown in the middle of Figure 7, with curves at
1-, 2- and 3-loop order for the potential difference. 
Although progress was made in studies of important finite volume effects, more work is needed to bring the systematics
under full control. It is worth noting that $V(R)$ and its slope tend to rise at large R-values with increasing spatial volumes.
In the current state of the analysis the string tension and the fast running coupling
are consistent with the $\rm{\chi SB}$ hypothesis and do not support the conformal one.

\subsection{Finite temperature transition}
We present some preliminary results from our more extended studies of the finite temperature
transition. If the ground state of the model has ${\rm \chi SB}$, a phase transition is expected
at some finite temperature in the chiral limit of massless fermions. This phase transition 
is expected to restore the chiral symmetry. If arguments based on universality, as implemented
in a model framework of  flavor dependence in the effective $\Phi^4$-theory description, were 
robust as advocated~\cite{Pisarski:1983ms},  the transition would be found to be of first order. 
This is not entirely clear and warrants careful continued investigations.
\begin{figure}[!ht]
\begin{center}
\includegraphics[width=0.35\textwidth]{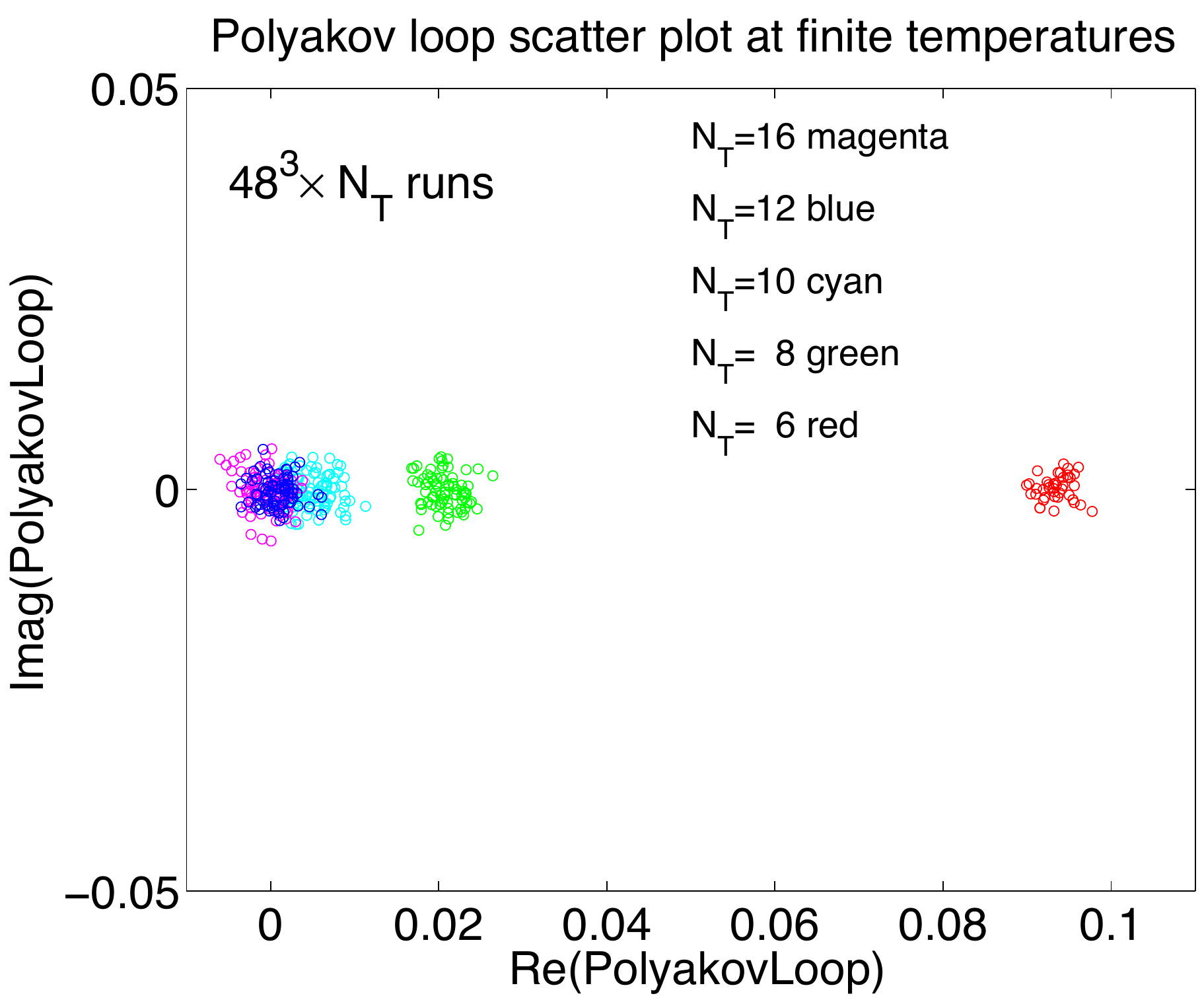}
\caption{{\footnotesize The scatter plot of the Polyakov loop in the time direction as the temperature is
varied using lattice sizes in a sequence of lattice sizes $48^3\times N_T$ at fixed $m=0.01$ with $N_T$ 
varied in the $N_T=6,8,10,12,16$ range.}}
\end{center}
\label{ploop}
\vskip -0.1in
\end{figure} 

On our largest lattice, at fixed $m=0.01$ and $\beta=2.2$, the temperature was varied through an $N_T$ sequence while the scatter plot of the Polyakov
loop was monitored along the euclidean time direction in each run. A clear sudden transition is observed in the $N_T=6-10$ region
where  the Polyakov loop
distribution jumps from the origin to a scatter plot with non-vanishing real part. 
It would be more difficult to reconcile this jump, as shown in Figure 8, with conformal
behavior in the zero temperature bulk phase.

Although we have results at other gauge couplings, and a variety of
fermion masses as the spatial volumes and the temperatures were varied, all consistent with a finite temperature transition,
caution is necessary before firm conclusions can be reached. 
Confirming the existence of the ${\rm \chi SB}$ phase transition will require the $m\rightarrow 0$  limits 
of  $\langle \bar{\psi}\psi\rangle$ and the Polyakov loop distribution.
The chiral condensate is a good order parameter for the transition. The Polyakov loop, like in QCD, could detect deconfinement
in the transition with well-known and somewhat problematic interpretation issues.

\section{Testing the hypothesis of conformal chiral symmetry}

\begin{figure*}[!htpb]
\begin{center}
\begin{tabular}{cc}
\includegraphics[height=5cm]{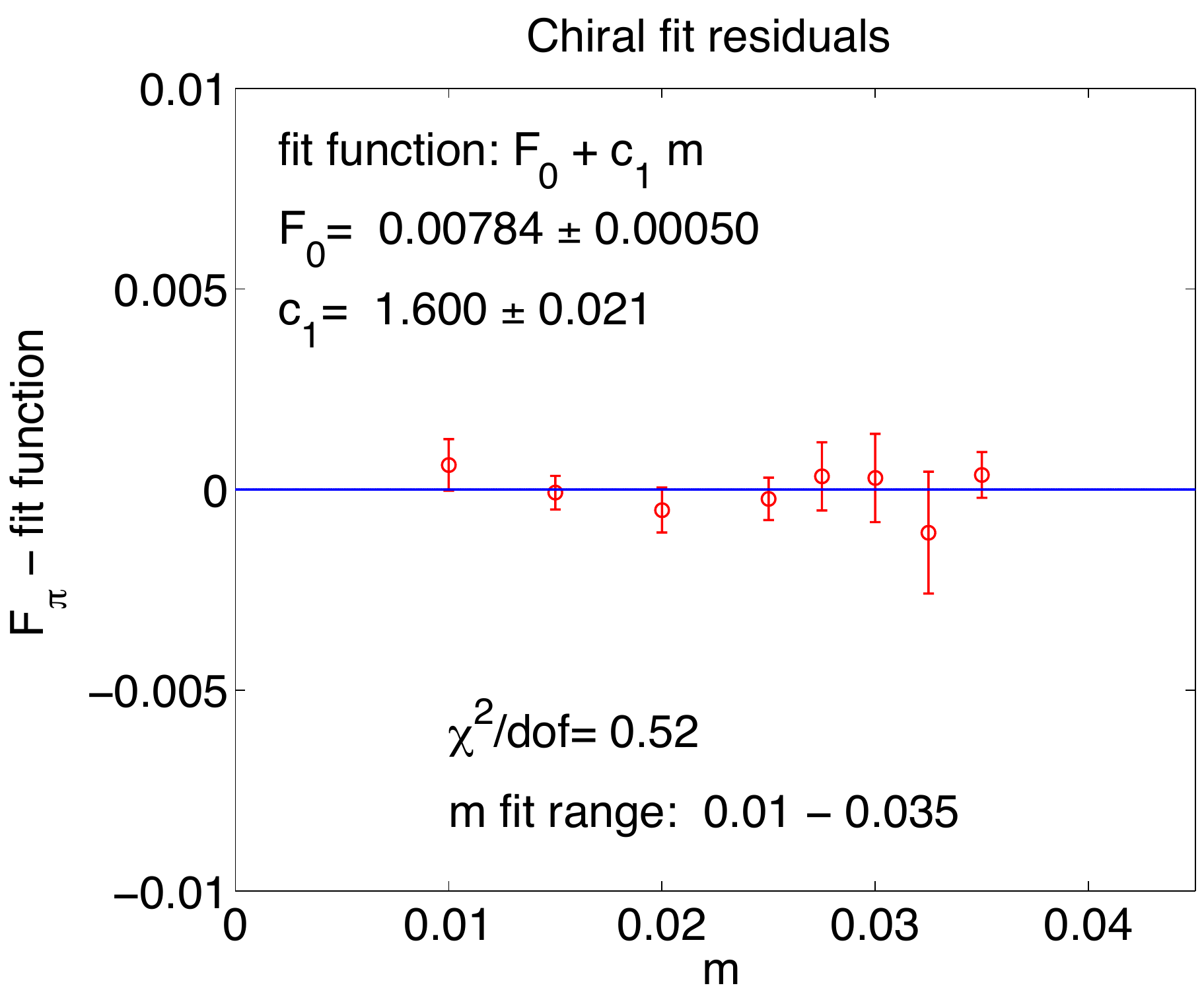}&
\includegraphics[height=4.9cm]{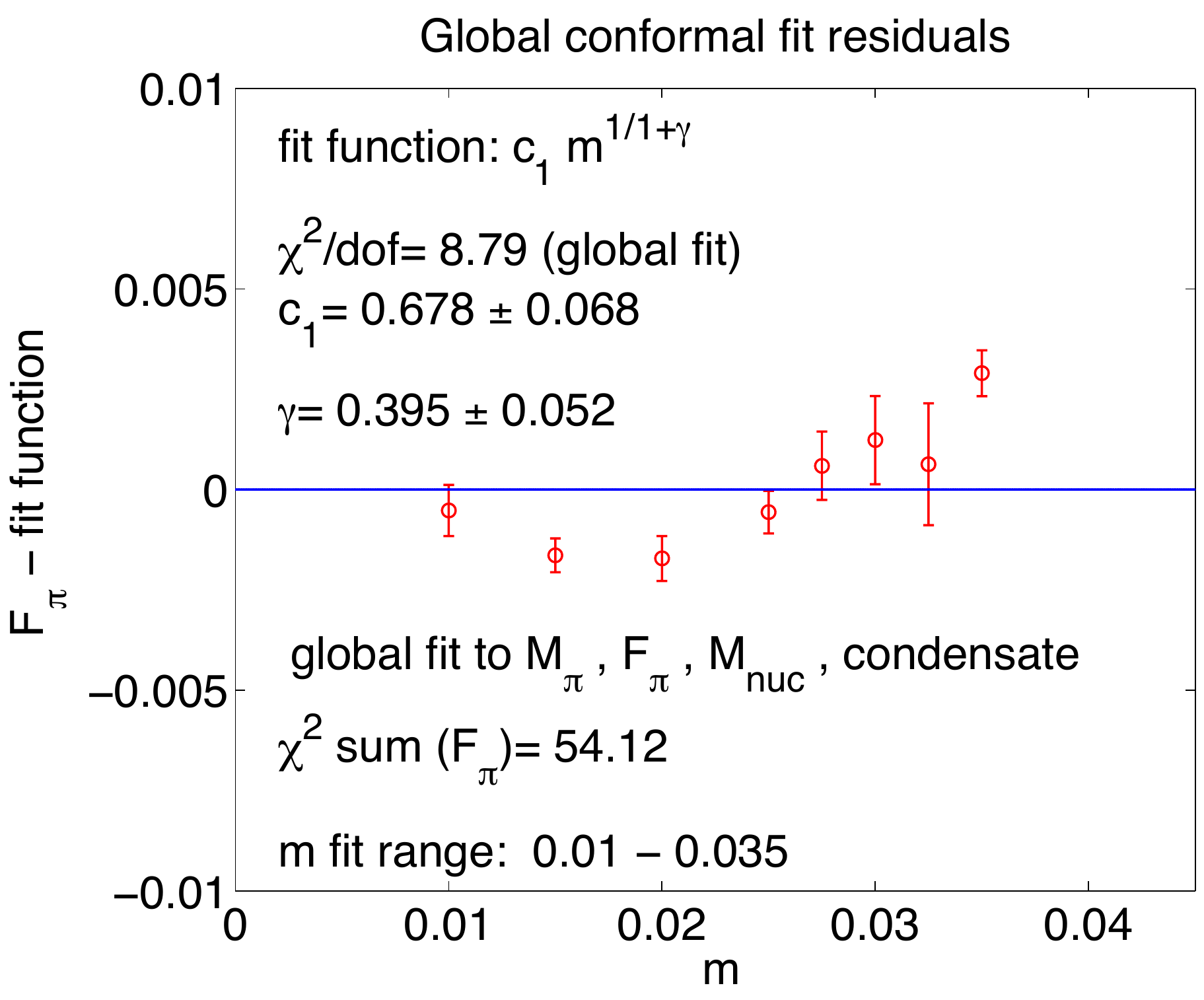}\\
\includegraphics[height=5.3cm]{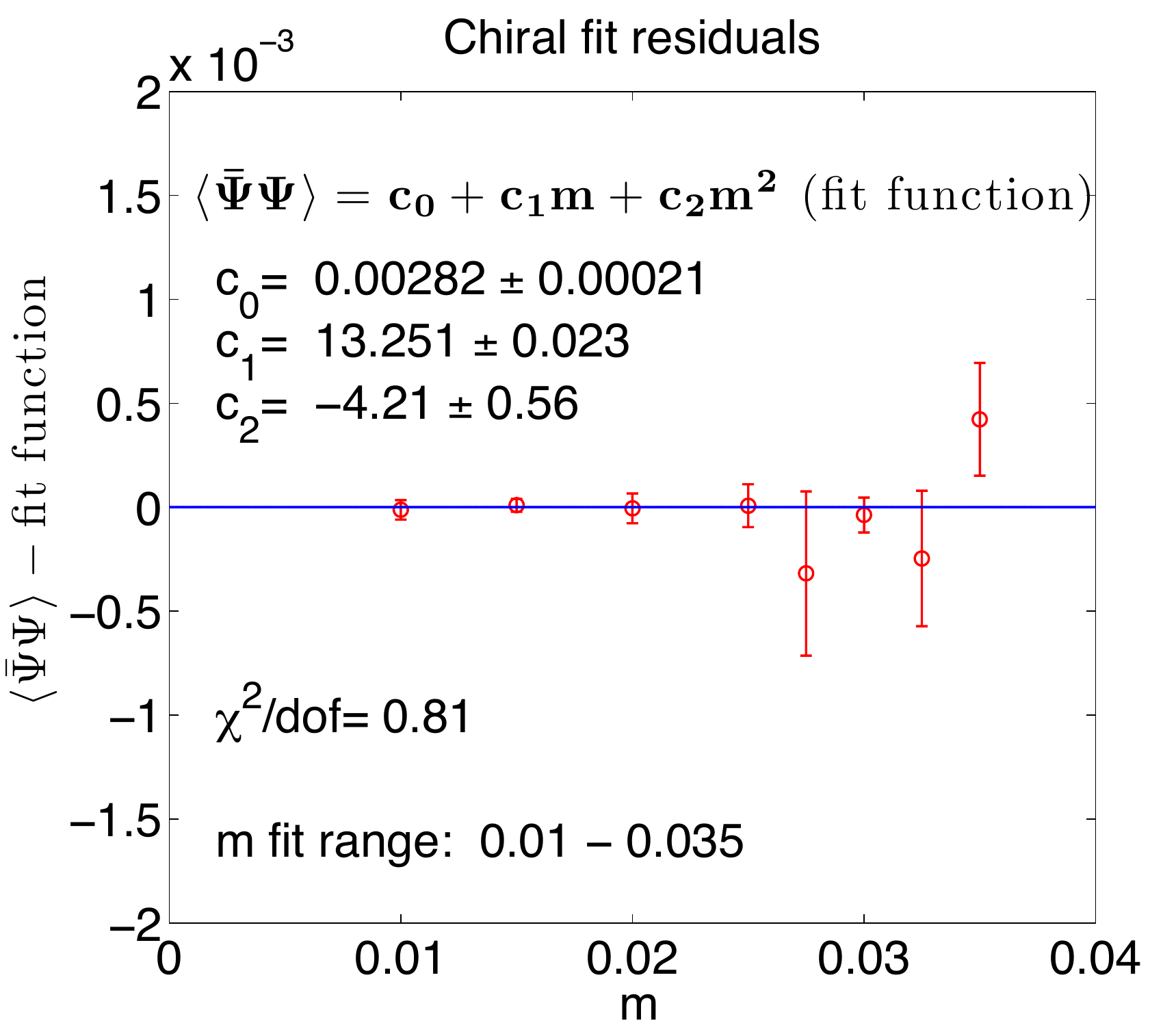}&
\includegraphics[height=5.5cm]{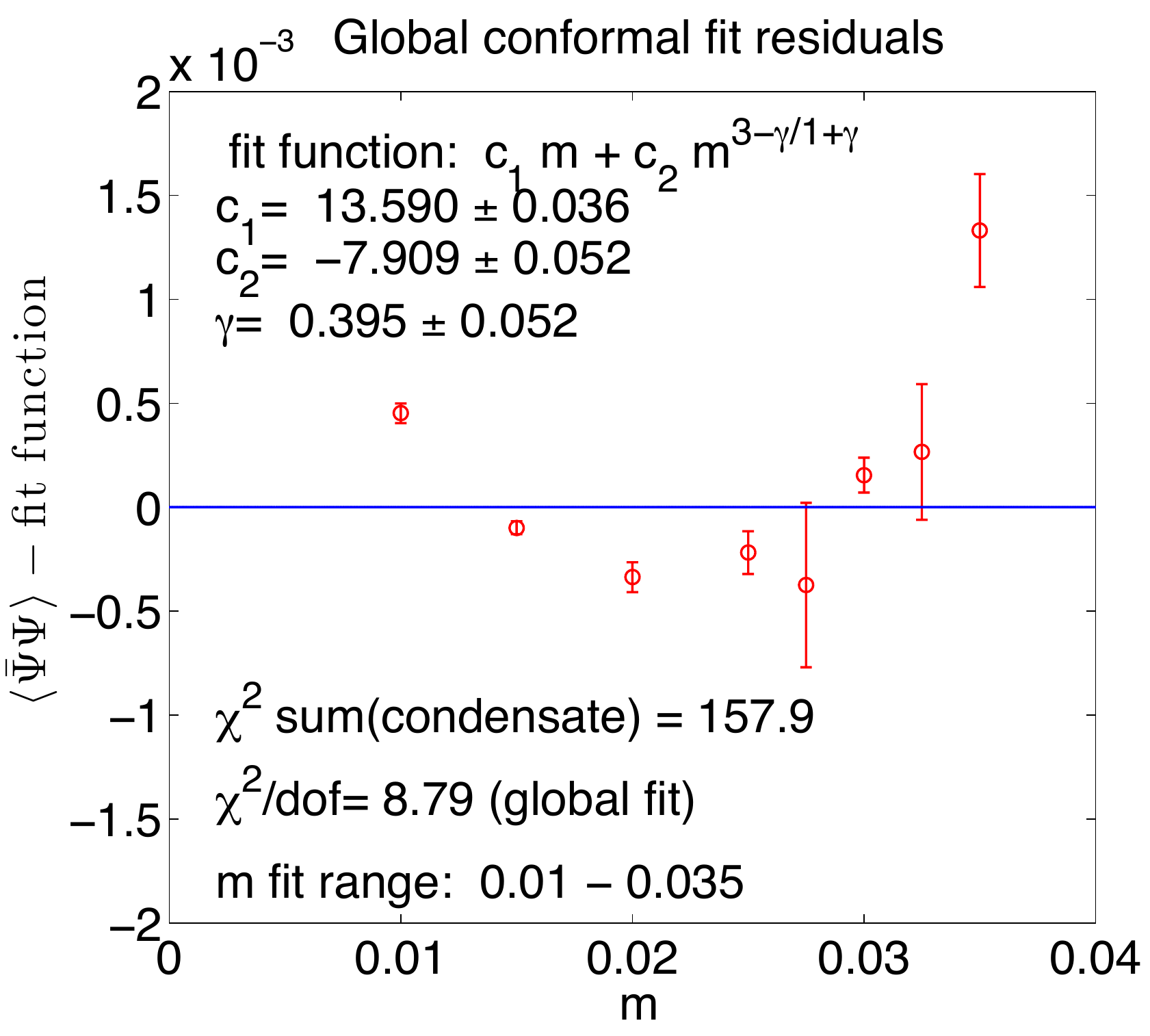}
\end{tabular}
\caption{{\footnotesize The $N_f=12$ chiral and conformal simultaneous fits in four channels are displayed for 
comparison in two select cases.}}
\end{center}
\label{fig:ConformNf12}
%\vskip -0.2in
\end{figure*}

The simulation results we presented  for twelve fermions in the fundamental
representation of the SU(3) color gauge group  favor the chiral symmetry
breaking hypothesis. The pion state is consistent with a vanishing mass in the chiral limit
and easy to fit with a simple quadratic function of the fermion mass. The non-Goldstone pion spectrum 
shows very little taste breaking at $\beta=2.2$ and the small splittings are consistent with expectations
for staggered fermions with stout smearing. The SO(4) degeneracies and splittings appear to follow the
pattern of QCD although the fermion mass dependence is significantly different. The fundamental 
scale-setting parameter $F$ of chiral symmetry breaking is finite in the chiral limit.

A non-vanishing chiral condensate is found in the chiral limit which is in the  ballpark of the GMOR relation as 
suggested by the small value of $F$. 
We find a consistent, non-vanishing chiral limit for the subtracted chiral condensate, with the dominant linear 
UV-contribution removed.
The nucleon states, 
the Higgs ($f_0$) meson, the $\rho$ meson and $A_1$ meson extrapolate to non-vanishing masses in 
the chiral limit and considerable splits of some of the parity partner states persist at very low fermion masses close to the
chiral limit. There seems to be an effective string tension indicating confinement-like behavior below the string-breaking scale
and the running coupling has 
not shown  signs of a fixed point slowdown.  In addition, there seems to be a rapid finite
temperature transition whose nature is unclear but hardly favors a conformal bulk phase. 
Our results are consistent with results reported in ~\cite{Jin:2009mc} but disagree with the chiral analysis
of ~\cite{Deuzeman:2009mh} and do not support the infrared fixed point reported
in~\cite{Appelquist:2007hu}.

But is it possible that we mislead ourselves with the $\chi{\rm SB}$ interpretation? Can we interpret the results as
conformal chiral symmetry? To decide this question, a fairly stringent test is possible. With the conformal hypothesis
the mass dependence 
of all physical states is controlled by the anomalous dimension $\gamma$  for small fermion masses~\cite{DelDebbio:2010hx}. Each
hadron and $F_\pi$ should scale as  $M_\pi \approx m^{1/y_m}$ and $F_\pi \approx m^{1/y_m}$ for small $m$ where 
$y_m=1+\gamma$. For small enough $m$ the value of $\gamma$ should be interpreted 
as $\gamma^{*}$ at the infrared fixed point. The chiral condensate is expected to have the behavior 
$\langle \bar{\psi}\psi\rangle \approx c\cdot m +  m^{\frac{3-\gamma}{1+\gamma}}$ when $m\rightarrow 0$.
We selected various subsets of states for a combined fit with universal critical exponent $\gamma$. We also fitted all measured
states combined. 
Applying the
conformal hypothesis to the chiral condensate, to $F$, to the pion state, and to the stable nucleon state collectively yields a total 
$\chi^2=229$ for 26 degrees of freedom with $\chi^2/{\rm dof}=8.79$. This indicates a very low 
level of confidence in the hypothesis. The  $\chi{\rm SB}$ hypothesis gives  $\chi^2/{\rm dof}=1.22$ for the same set of states. 
This was the result quoted in Section 1. 
The chiral and conformal fits for two of the four fitted states with the quoted global results are shown in Figure 9.
Applying the global analysis to all states we measured, the contrasting behavior is less pronounced but still significant. 
The results disfavoring the conformal hypothesis are significant. More work is needed for higher accuracy and full control of the systematics,
yet it is worth noting that as the volumes are increased at the lower quark masses, the results for $F_\pi$
and $\langle \bar{\psi}\psi\rangle$  will increase or remain the same; this does not bode well for the conformal picture.

\section{Conclusions and outlook}

We reported new results for a frequently discussed gauge theory with twelve fermion flavors  
in the fundamental  representation of the SU(3) color gauge group. Our results
favor with a significant level of confidence the ${\rm\chi SB}$ scenario but
close to the conformal window several features of this gauge theory clearly differ from QCD. 
We find a large $B/F$ ratio which is often interpreted as strong chiral condensate enhancement. This could
be related to the fermion mass dependence of the Goldstone pion and the non-Goldstone pion spectra  which is 
different from what is observed in QCD with staggered fermions. In comparison with QCD,
we also observe significantly smaller mass splittings between parity partners in several hadron channels. 
The near degeneracy of parity partner states is expected to lead to an S-parameter quite different from what
was projected when QCD was scaled up by the number of flavors~\cite{Peskin:1991sw}.
To gain more confidence in our ${\rm\chi SB}$ analysis, it is important to push deeper into the 
chiral regime and closer to the continuum limit than the analysis  reported earlier~\cite{Deuzeman:2009mh}.
Only future work with high precision simulations will be able to explain whether this is the source of
our qualitatively different findings.

The infrared fixed point (IRFP) of the gauge coupling reported earlier ~\cite{Appelquist:2007hu} 
is also in disagreement with the
picture we presented. It seems to be very difficult to differentiate between an IRFP and a 
slowly walking gauge coupling close to the
conformal window as also indicated with MCRG studies of the  model~\cite{Hasenfratz:2010fi}.
It would be interesting to establish in a quantitative analysis using the Schr\"odinger Functional method
of ~\cite{Appelquist:2007hu} and the MCRG method of ~\cite{Hasenfratz:2010fi} the difference in the level
of confidence between the two mutually exclusive hypotheses of IRFP or walking gauge coupling. This was precisely
the main thrust of our work in comparing the ${\rm\chi SB}$ hypothesis with the conformal hypothesis.

We plan the continued investigation of the running gauge coupling reported in Section 2 with 
the ambitious goal of differentiating 
between the IRFP and walking gauge coupling scenarios. 
If a walking gauge coupling $g^2_w$ can be established  approximating
IRFP  behavior over some extended scale, it will
be important to determine the related mass anomalous dimension $\gamma(g^2_w)$ and the S-parameter as model
examples for BSM applications.
%If a walking gauge coupling $g^2_w$ can be established with approximate
%IRFP  behavior over some extended scale, $g^2_w\approx \!g^2_{IRFP} $, 

\section*{Acknowledgments}
The simulations were performed using computational resources
  at Fermilab and JLab, under the auspices of USQCD and SciDAC, from the
  Teragrid structure and at Wuppertal. We are grateful to Kalman Szabo, Sandor Katz, and Stefan Krieg
for helping us in using the Wuppertal RHMC code. 
Simulation on GPU clusters were facilitated by the CUDA ports of \cite{Egri:2006zm}.
This research is supported by
the NSF under grants 0704171 and 0970137, by the DOE under grants
 DOE-FG03-97ER40546, DOE-FG-02-97ER25308, by the DFG under grant FO
 502/1 and by SFB-TR/55, and the EU Framework Programme 7 grant
 (FP7/2007-2013)/ERC No 208740.
D.N. would like to thank the Aspen Center for Physics for invitation to the 2010 BSM summer program.
C.R.S. is indebted to the Theory Group at CERN for hospitality at the LGT 2010 workshop.

\end{document}